\documentclass[useAMS,usenatbib]{mn2e}
\usepackage{graphicx}
\usepackage{hyperref}
\hypersetup{backref,pdfpagemode=FullScreen,colorlinks=true,citecolor=blue} 

\defcitealias{simard11}{S11}
\defcitealias{mendel12}{M12}

\def\msol{\hbox{\kern 0.20em $M_\odot$}}
\def\lsol{\hbox{\kern 0.20em $L_\odot$}}
\def\rsol{\hbox{\kern 0.20em $R_\odot$}}
\def\sr{\hbox{\kern 0.20em sr}}
\def\srmu{\hbox{\kern 0.20em sr$^{-1}$}}

\def\g{\hbox{\kern 0.20em g}}
\def\gmu{\hbox{\kern 0.20em g$^{-1}$}}
\def\kg{\hbox{\kern 0.20em kg}}
\def\pc{\hbox{\kern 0.20em pc}}

\def\mum{\hbox{\kern 0.20em $\mu$m}}
\def\mumd{\hbox{\kern 0.20em $\mu$m$^{-2}$}}
\def\cm{\hbox{\kern 0.20em cm}}
\def\m{\hbox{\kern 0.20em m}}
\def\km{\hbox{\kern 0.20em km}}
\def\nm{\hbox{\kern 0.20em nm}}

\def\s{\hbox{\kern 0.20em s}}
\def\h{\hbox{\kern 0.20em h}}
\def\sec{\hbox{\kern 0.20em sec}}
\def\min{\hbox {\kern 0.20em min}}
\def\smu{\hbox{\kern 0.20em s$^{-1}$}}
\def\smd{\hbox{\kern 0.20em s$^{-2}$}}
\def\an{\hbox{\kern 0.20em an}}
\def\anmu{\hbox{\kern 0.20em an$^{-1}$}}
\def\deg{\hbox{\kern 0.20em $^{\rm o}$}}
\def\yr{\hbox{\kern 0.20em yr}}
\def\yrmu{\hbox{\kern 0.20em yr$^{-1}$}}
\def\Myr{\hbox{\kern 0.20em Myr}}
\def\Mymu{\hbox{\kern 0.20em Myr$^{-1}$}}
\def\K{\hbox{\kern 0.20em K}}
\def\pcmu{\hbox{\kern 0.20em pc$^{-1}$}}
\def\pcmd{\hbox{\kern 0.20em pc$^{-2}$}}
\def\pcmt{\hbox{\kern 0.20em pc$^{-3}$}}
\def\kms{\hbox{\kern 0.20em km\kern 0.20em s$^{-1}$}}
\def\kmpd{\hbox{\kern 0.20em km$^{2}$}}
\def\kpc{\hbox{\kern 0.20em kpc}}
\def\cms{\hbox{\kern 0.20em cm\kern 0.20em s$^{-1}$}}
\def\erg{\hbox{\kern 0.20em erg}}
\def\ergs{\hbox{\kern 0.20em erg}}
\def\cmpd{\hbox{\kern 0.20em cm$^2$}}
\def\cmmd{\hbox{\kern 0.20em cm$^{-2}$}}
\def\cmms{\hbox{\kern 0.20em cm$^{-6}$}}
\def\cmpt{\hbox{\kern 0.20em cm$^3$}}
\def\cmmt{\hbox{\kern 0.20em cm$^{-3}$}}
\def\mpd{\hbox{\kern 0.20em m$^2$}}
\def\mmd{\hbox{\kern 0.20em m$^{-2}$}}
\def\mpt{\hbox{\kern 0.20em m$^3$}}
\def\mmt{\hbox{\kern 0.20em m$^{-3}$}}
\def\mujy{\hbox{\kern 0.20em $\mu$Jy}}
\def\mjy{\hbox{\kern 0.20em mJy}}
\def\Mj{\hbox{\kern 0.20em MJy}}
\def\jy{\hbox{\kern 0.20em Jy}}
\def\ghz{\hbox{\kern 0.20em GHz}}
\def\srmd{\hbox{\kern 0.20em sr$^{-1}$}}
\def \kms{km~$\rm{s}^{-1}$}

\def \mum{$\mu$m}

\def\G{\hbox{\kern 0.20em G}}

\def\h13cop{\hbox{H$^{13}$CO$^{+}$}}

\def\S+{\hbox{S{\small II}}}

\title[SMBH in the Hubble Sequence]{The Ubiquity of Supermassive
Black Holes in the Hubble Sequence}

\author[Marleau et al.]{Francine R.\ Marleau, Dominic Clancy and Matteo Bianconi\\
Institute of Astro and Particle Physics, University of Innsbruck, 6020 Innsbruck, Austria}

\begin{document}

\date{Accepted 8 August 2013}


\maketitle

\label{firstpage}

\begin{abstract}
We present the results of a study of a statistically significant
sample of galaxies which clearly demonstrate that supermassive black
holes are generically present in all morphological types. Our analysis
is based on the quantitative morphological classification of 1.12
million galaxies in the SDSS DR7 and on the detection of black hole
activity via two different methods, the first one based on their
X-ray/radio emission and the second one based on their mid-infrared
colors. The results of the first analysis confirm the correlation
between black hole and total stellar mass for 8 galaxies and includes
one galaxy classified as bulgeless. The results of our second
analysis, consisting of 15,991 galaxies, show that galaxies hosting a
supermassive black hole follow the same morphological distribution as
the general population of galaxies in the same redshift range. In
particular, the fraction of bulgeless galaxies, 1,450 galaxies or 9
percent, is found to be the same as in the general population.  We
also present the correlation between black hole and total stellar mass
for 6,247 of these galaxies. Importantly, whereas previous studies
were limited to primarily bulge-dominated systems, our study confirms
this relationship to all morphological types, in particular, to 530
bulgeless galaxies. Our results indicate that the true correlation
that exists for supermassive black holes and their host galaxies is
between the black hole mass and the total stellar mass of the galaxy
and hence, we conclude that the previous assumption that the black
hole mass is correlated with the bulge mass is only approximately
correct.
\end{abstract}


\section{Introduction}
\label{sec:introduction}

According to the standard picture, the structure we observe today in
our Universe formed as a consequence of the growth of fluctuations in
the primordial dark matter distribution which gravitationally
attracted gaseous baryonic matter that later settled into disk
galaxies \citep[e.g.][]{white78, white91}. Galaxy mergers within halos
subsequently transformed disks into bulges and the gas fell to the
center, triggering starbursts and feeding the rapid growth of black
holes (BH). These in turn responded by feeding energy back to the
surrounding gas \citep[e.g.][]{sander88, hopkins08}. This picture of
structure formation is based on observational work which, over the
past few decades, indicates that most -if not all- {\em massive}
galaxies with a spheroidal component have a supermassive black hole
(SMBH) at their center \citep[e.g.][]{kormendy95, magorrian98}. In
addition, estimates of the BH masses of these giant galaxies have been
found to tightly correlate with the spheroid/host luminosity (or
stellar mass) and also with the stellar velocity dispersion within the
host bulges \citep[e.g.][]{gebhardt00, ferrarese00}.

However, there is growing evidence that SMBH not only exist at the
centers of massive galaxies with a central bulge, but also at the
centers of galaxies with pseudo-bulges and that these systems also
exhibit a correlation between black hole mass and total stellar mass
\citep{jiang11, simmons13}. This supports the alternative view that
SMBH growth might be due to minor mergers and/or secular processes
\citep[e.g.][]{greene10, simmons13}. Although, as pointed out by
\citet{okamoto12}, the main channel of pseudo-bulge formation may in
fact not be secular disk evolution but rather rapid gas supply at
high-redshift.

Given the uncertainties concerning the current understanding of the
formation of bulges and pseudo-bulges in galaxies, the problem of
identifying the main process at play in the formation of SMBH at the
centers of galaxies with a spheroidal component remains a difficult
challenge at present. On the other hand, the recent discovery of a
SMBH at the center of Henize 2-10 \citep{reines11}, a dwarf starburst
galaxy lacking any substantial spheroidal component, suggests that a
merger is not essential to making a SMBH and that BH growth may after
all require substantial and ongoing secular processes, at least in
these galaxies. Indeed, recent theoretical developments on the
formation of the first massive BH show that they may form out of
gas-dynamical instability at low metallicity
\citep[e.g.][]{volonteri12}. Moreover, if SMBH are also generic to
bulgeless galaxies and dwarf galaxies, as indeed the current work will
present evidence to demonstrate they are, these galaxy types may offer
the best probe of SMBH formation and evolution in the local Universe.
In this regard, a SMBH hosted by a bulgeless galaxy and/or dwarf
galaxy is more likely ``pristine'', as it is unlikely that it has
experienced mergers and dynamical interactions but has rather
undergone a quieter evolution. Moreover, pure-disk galaxies are far
from rare: locally, only four of the nineteen giant galaxies are
ellipticals or have classical bulges \citep{kormendy10}. While in the
Sloan Digital Sky Survey \citep[SDSS;][]{york00}, based on the
quantitative morphological analysis of 1.12 million galaxies in the
Data Release Seven \citep[DR7;][]{abazajian09}, the fraction of
galaxies with an optical S\'ersic profile index $n_g < 1.5$ and bulge
fraction $B/T < 0.1$, i.e.\ nearly bulgeless, is $\sim$14\%
\citep[hereafter \citetalias{simard11}]{simard11}. However, despite
the large fraction of bulgeless galaxies and their clear importance in
challenging our current understanding of galaxy formation and
evolution, a systematic survey of SMBHs in bulgeless galaxies has
never been undertaken, though this is in part due to the lack of a
well-defined sample. In this paper, we attempt to rectify this
situation and substantially extend the status of the present knowledge
of SMBH and their host galaxies by presenting the first systematic
survey of SMBHs in galaxies of all morphological types,
i.e.\ including those classified as bulgeless galaxies, based on the
quantitative morphological analysis of 1.12 million galaxies in the
SDSS DR7 \citepalias{simard11}.

In this work we report the results of two separate studies which
explore the nature of SMBH and their host galaxies, the first based on
an X-ray/radio selected sample and the second on an infrared selected
sample.  The structure of our paper is as follows. In
Section~\ref{sec:quantitative}, we describe the method that we used to
select our sample of morphologically classified galaxies. In
Section~\ref{sec:xray}, we present the selection and analysis of our
X-ray/radio selected sample of active SMBH, while in
Section~\ref{sec:infrared}, we present the selection and analysis of
our infrared selected sample of active SMBH. In
Section~\ref{sec:summary}, we summarize our results. Considered
together, we believe these two studies are complementary in nature, in
that while our first analysis of an X-ray selected sample provides a
well-tested and robust estimate of black hole mass but with a small
statistic, our second analysis of an infrared selected sample provides
estimates on the black hole mass, parametrised by their accretion
efficiency, but with a large statistic.

\begin{figure}
\centerline{
\includegraphics[width=240pt,height=240pt,angle=0]{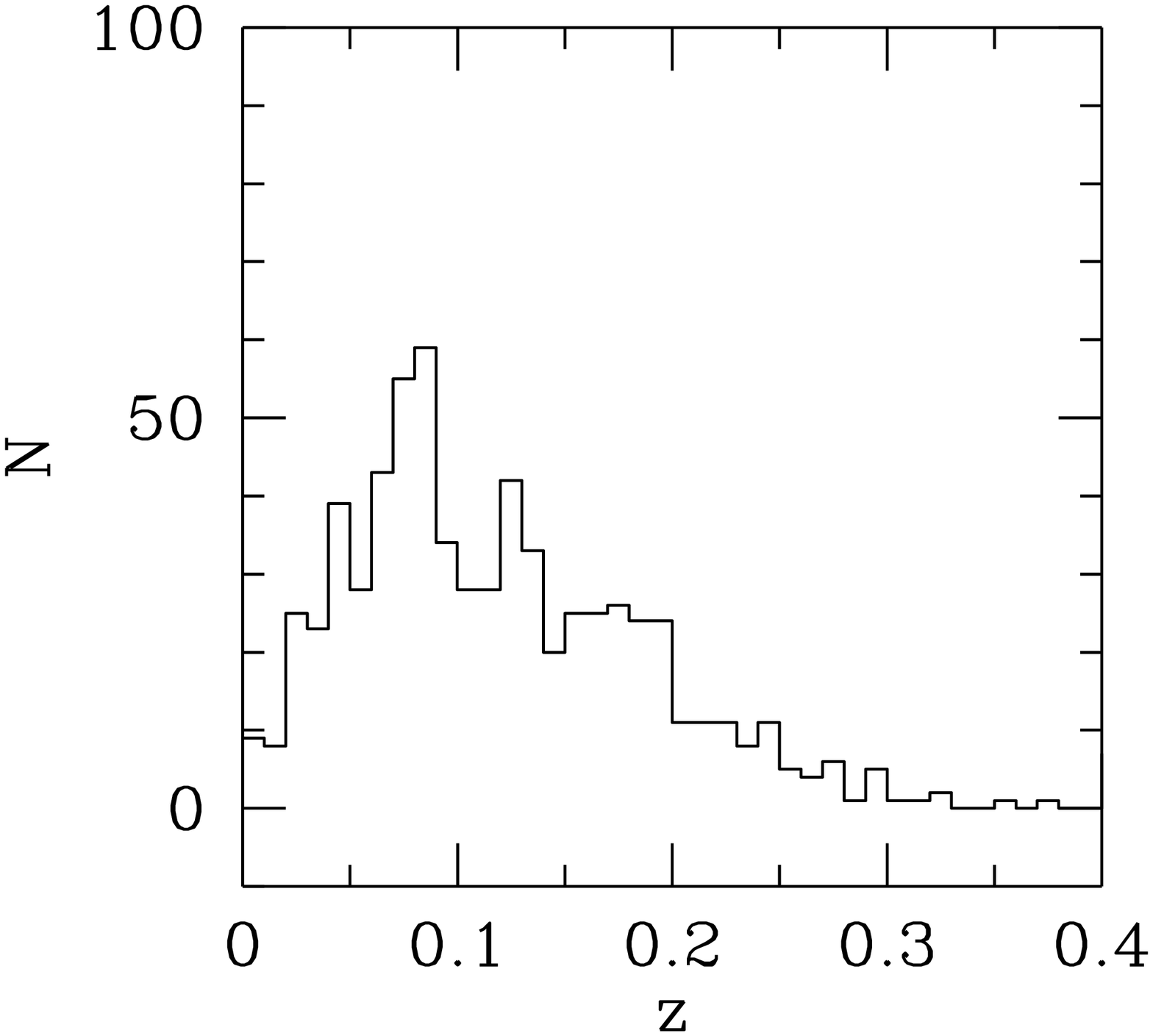}
}
\centerline{
\includegraphics[width=240pt,height=240pt,angle=0]{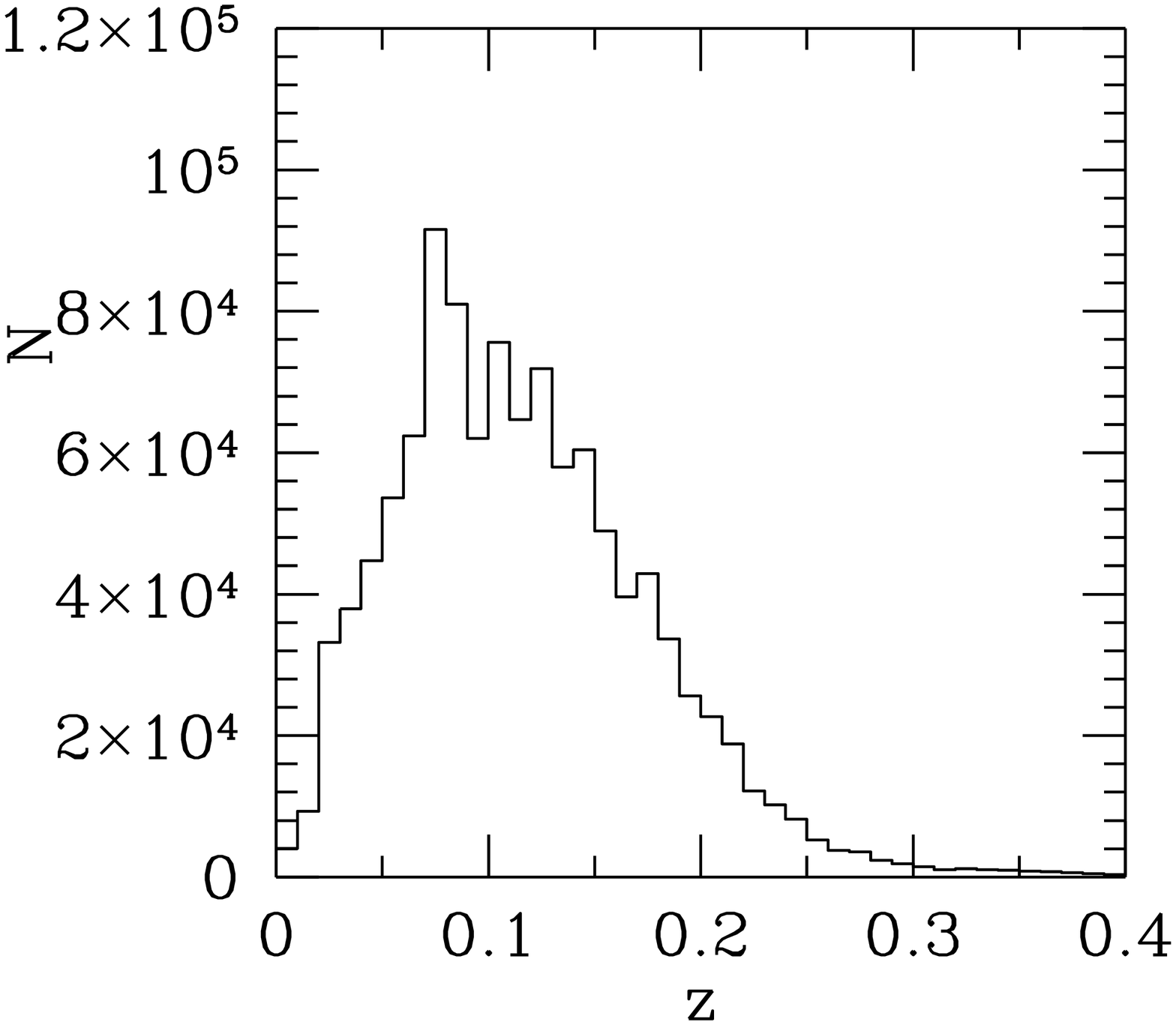}
}
\caption{\label{fig:hist_z_Xray} {\it Top}: Redshift distribution of the 688 
 galaxies in the CSC-SDSS catalog with structural parameters. 
{\it Bottom}: Redshift distribution of the SDSS
  DR7 galaxies in the \citetalias{simard11} morphological catalog. \vspace{0.2cm}}
\end{figure}

\begin{figure*}
\centerline{
\includegraphics[width=160pt,height=160pt,angle=0]{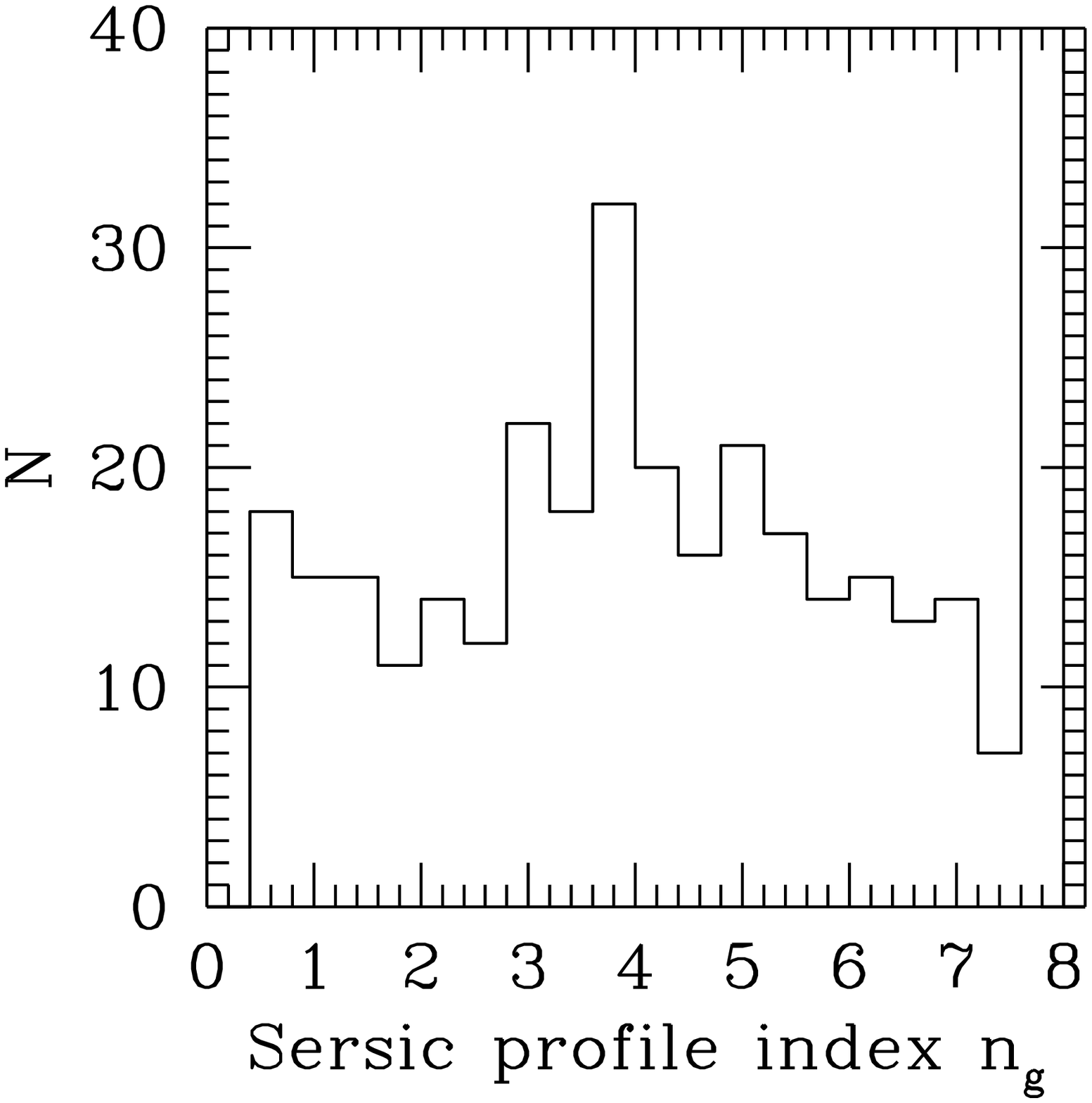}\includegraphics[width=160pt,height=160pt,angle=0]{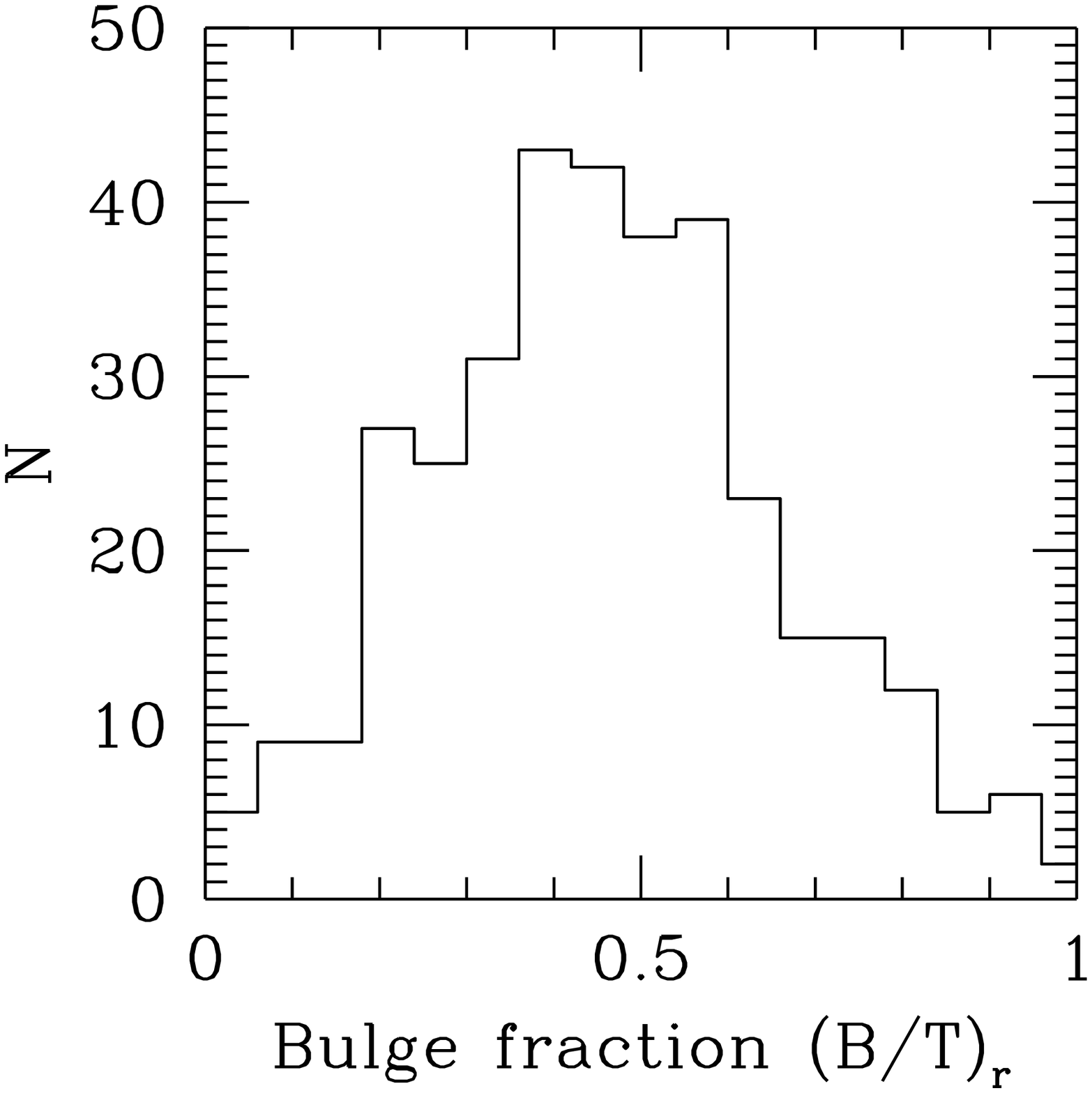}\includegraphics[width=160pt,height=160pt,angle=0]{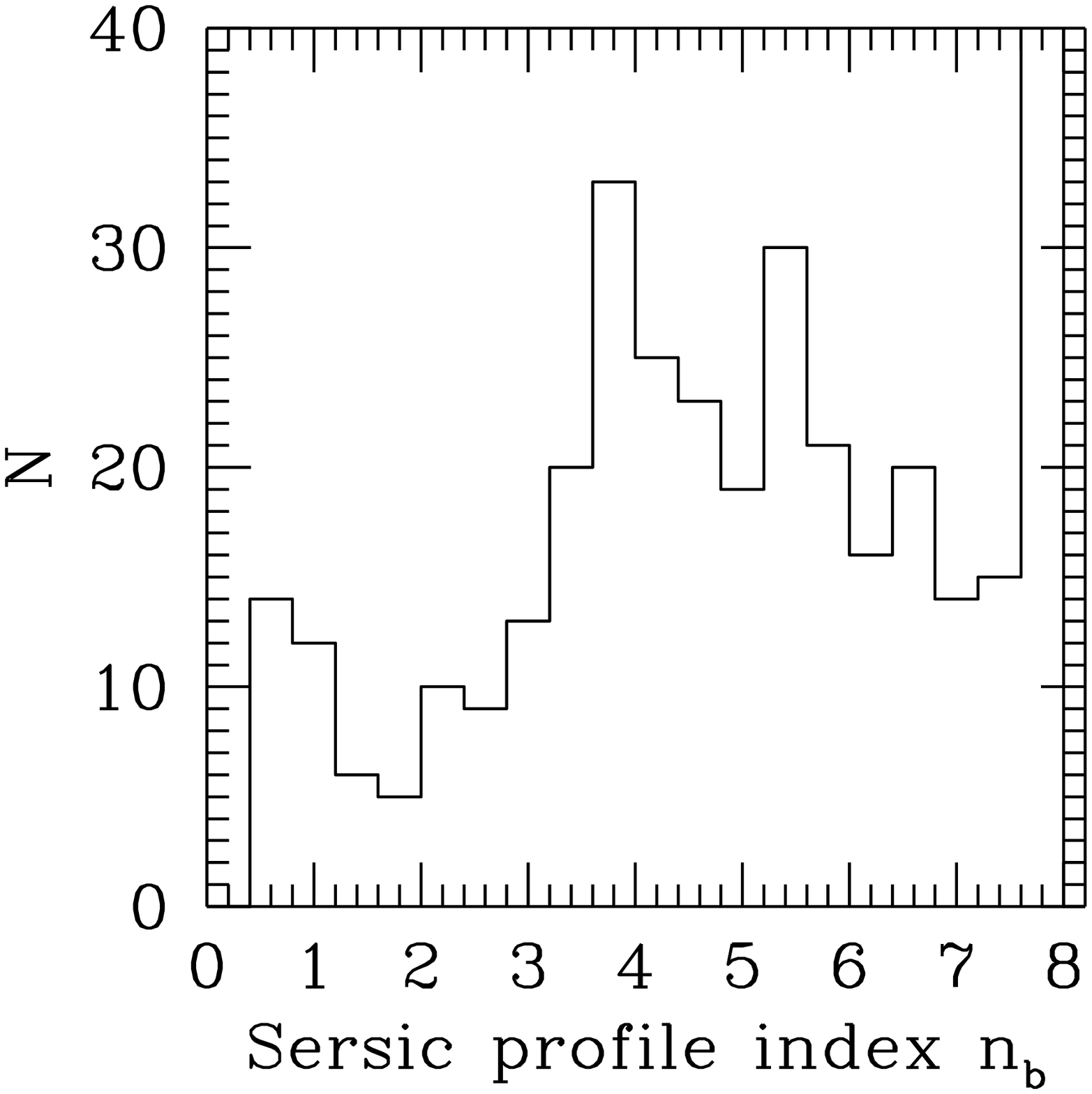}
}
\centerline{
\includegraphics[width=160pt,height=160pt,angle=0]{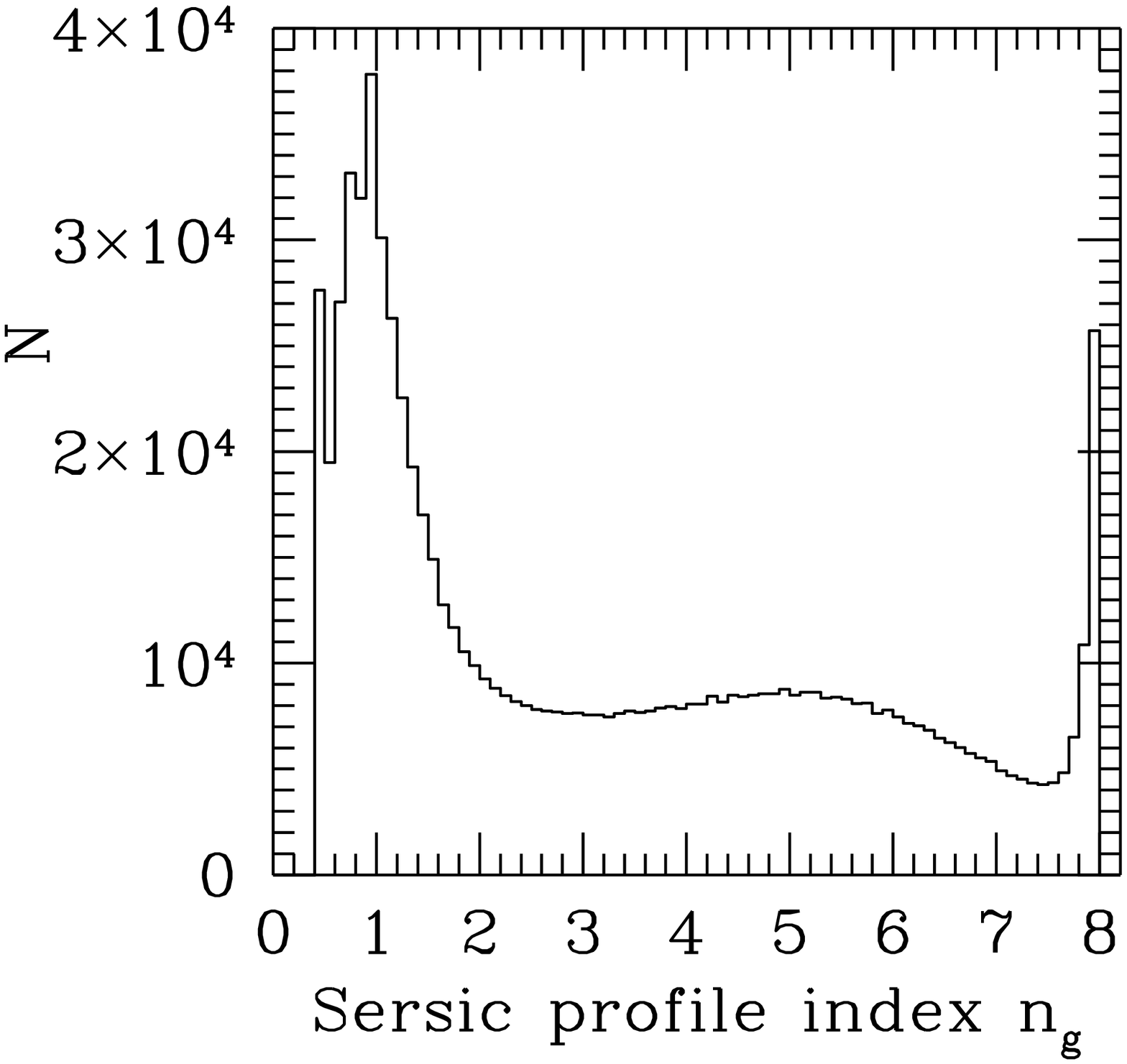}\includegraphics[width=160pt,height=160pt,angle=0]{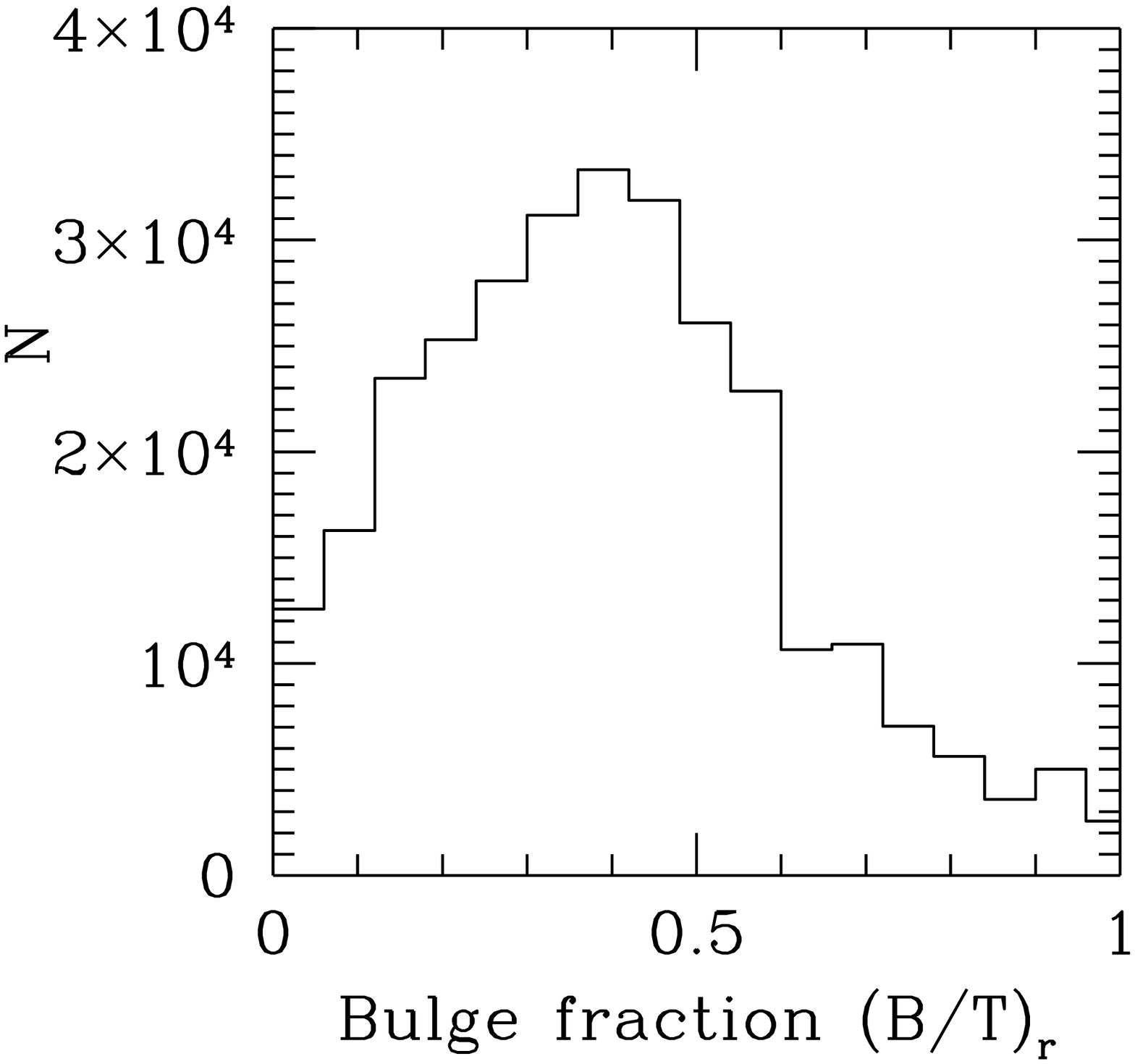}\includegraphics[width=160pt,height=160pt,angle=0]{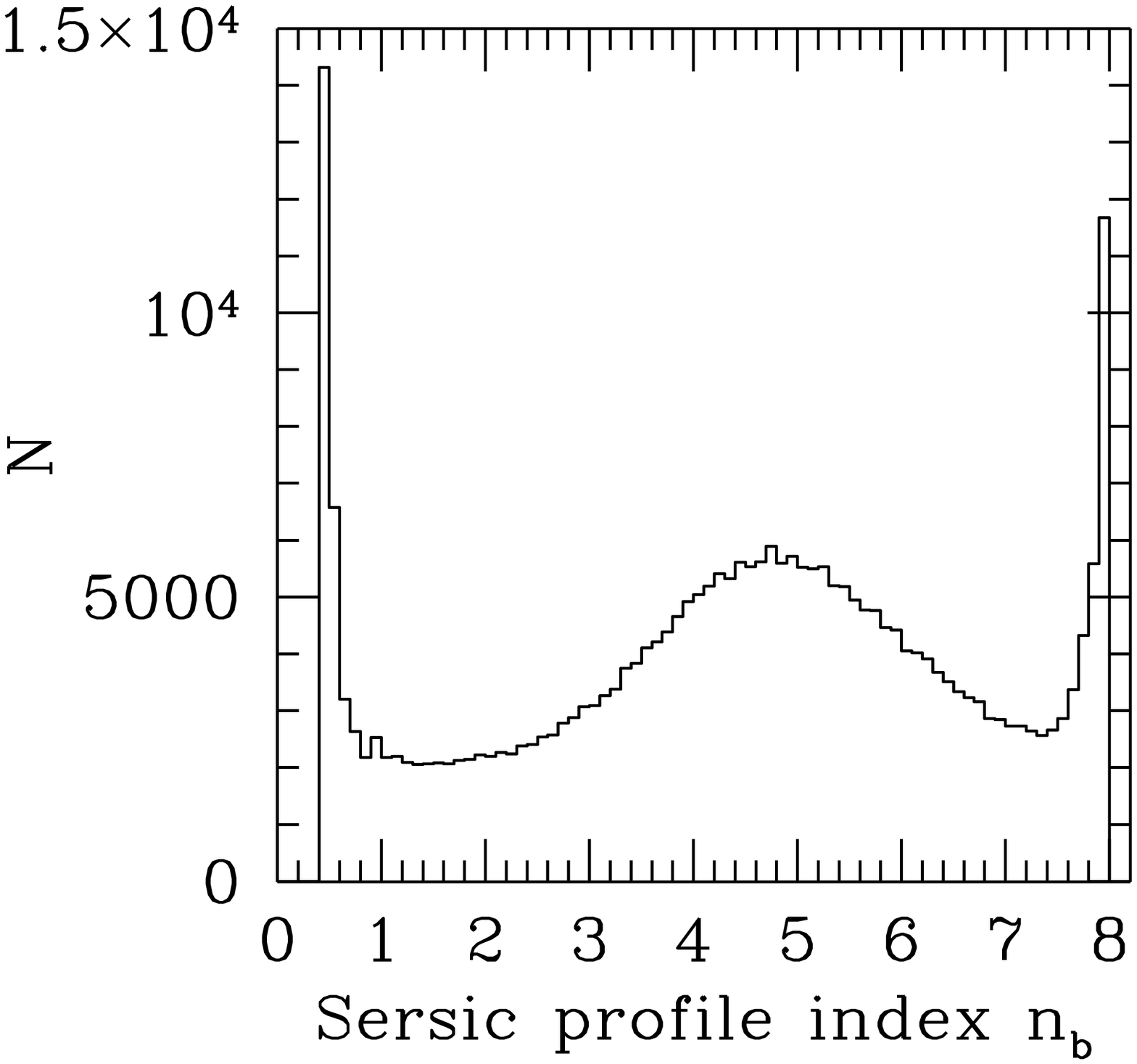}
}
\caption{\label{fig:hist_ngbtr_Xray} {\it Top row}: Distribution of the S\'ersic
  index $n_g$ ({\it left}), the bulge fraction $(B/T)_r$ ({\it
    middle}) and the bulge S\'ersic index $n_b$ ({\it right}) of the
  galaxies in the CSC-SDSS catalog with structural parameters. 
{\it Bottom row}: Distribution of the S\'ersic profile
  index $n_g$ ({\it left}), the bulge fraction $(B/T)_r$ ({\it
    middle}) and the bulge S\'ersic index $n_b$ ({\it right}) of the SDSS
  DR7 galaxies in the \citetalias{simard11} morphological catalog.
\vspace{0.2cm}}
\end{figure*}

\section{Quantitative Morphological Parameters}
\label{sec:quantitative}

As the foundation of all our work we used the quantitative
morphological classification of \citetalias{simard11} to extract
structural parameters for 1.12 million galaxies. In this work, the
sources that were fitted were selected from the Legacy area with $14.0
\leq m_{r} \leq 18.0$, where $m_{r}$ is the $r$-band Petrosian
magnitude corrected for Galactic extinction according to the
extinction values given in the SDSS database, and morphological type
Type = 3, i.e.\ sources classified as galaxies. In addition to these
two criteria, the sum of the flags DEBLENDED\_AS\_PSF and SATURATED
were required to be zero to eliminate objects that were found to be
unresolved children of their parents as well as saturated objects. The
total number of objects in the \citetalias{simard11} catalog is
1,123,718.

Galaxy structural parameters obtained by \citetalias{simard11} were
measured from bulge + disk decompositions performed using version 3.2
of the GIM2D software package \citep{simard02}. In addition to the
canonical de Vaucouleurs bulge (S\'ersic index $n_b = 4$) +
exponential disk fitting model \citepalias[Table~1]{simard11}, two
other fitting models were used. The first additional model was a free
$n_b$ bulge + disk model, with $n_b$ allowed to vary from 0.5 to 8
\citepalias[Table~2]{simard11}, and the second one was a single
component pure S\'ersic model, with $n_g$ also allowed to vary from
0.5 to 8 \citepalias[Table~3]{simard11}. The appropriate model was 
selected using the F-test probability $P_{pS}$ and the uncertainties
associated with the best fit parameters were estimated from
the 99\% confidence limits \citep[Table 1-3, \citetalias{simard11};][]{marleau98}.

In order to classify a galaxy as bulgeless, we required an optical
S\'ersic profile index $n_g < 1.5$ and a bulge fraction $B/T <
0.1$. For the whole \citetalias{simard11} catalog, this yielded a list
of 88,594 galaxies, corresponding to about 14\% of the total sample.
As shown by \citet{kelvin13}, the ability of the quantitative code to
recover bulges is limited by the resolution of the images, in this
case 1.2 arcsec. For a typical bulge of size 3~kpc
\citepalias{simard11}, quantitative methods can recover bulges in the
SDSS images to a redshift of about 0.06. Assuming a size distribution
that is Gaussian, we would expect to be able to correctly identify
larger bulges at higher redshift and smaller bulges to lower
redshift. For our bulgeless sample, only 8\% of these galaxies fall
below $z = 0.06$. From this we conclude that the number of bulgeless
galaxies we have identified may be an overestimate as, with higher
resolution images, we are likely to find bulges in the large fraction
of these galaxies at $z > 0.06$.

\section{X-ray SMBH Sample Selection}
\label{sec:xray}

\subsection{CSC-SDSS Cross-Match Catalog}
\label{sec:xray.1} 

In our first analysis, which we now describe, we investigated the
properties of a sample of galaxies containing AGN selected via their
X-ray detection. In order to achieve this purpose we utilized the
Chandra Source Catalog/Sloan Digital Sky Survey
\citep[CSC-SDSS;][]{evans10} cross-match
catalogue\footnote{http://cxc.harvard.edu/cgi-gen/cda/CSC-SDSSxmatch.html}
to select active black holes via their X-ray emission. The X-ray color
(or hardness ratio), often used to identify AGN, was left
unconstrained in this selection process in order to avoid filtering
out some AGN candidates, as, for example, many of the brightest X-ray
sources are identified as AGN candidates by the Wide-field Infrared
Survey Explorer \citep[WISE;][]{wright10}, regardless of their
hardness ratio \citep{wang04, stern12}. In total, 8,997 galaxies in
the SDSS were cross-matched to a Chandra X-ray detection. Of these, we
subsequently identified 688 as having been assigned structural
parameters in the catalog of \citetalias{simard11}. Finally, after
applying our selection criteria of $n_g < 1.5$ and $B/T < 0.1$, we
were left with a sample of 26 galaxies classed as bulgeless.

\subsection{Redshift Distribution and Structural Parameters}
\label{sec:xray.2}

In order to gain some insight into the nature of our of X-ray selected
sample of 688 galaxies with structural parameters, we examined their
redshift and structural parameter distributions and compared them to
the general population. The redshift distribution of the X-ray
selected sample has been plotted in Figure~\ref{fig:hist_z_Xray},
while the distributions in S\'ersic index, bulge fraction and bulge
S\'ersic index are depicted in Figure~\ref{fig:hist_ngbtr_Xray}. The
S\'ersic profile index was plotted only for galaxies better fitted
with a S\'ersic model, i.e.\ with an F-test probability $P_{pS} >
0.32$ \citepalias{simard11}, whereas both the bulge fraction and the
bulge S\'ersic index were plotted only for galaxies better fitted with
a bulge + disk components model, in this case with $P_{pS} \leq
0.32$. The shape of the redshift distribution was found to be similar
to that of \citetalias{simard11} (see Figure~\ref{fig:hist_z_Xray}),
with a peak at $z \sim 0.1$, indicating no obvious redshift bias in
the X-ray sample.  However, the distribution in S\'ersic profile was
found to differ, being flatter at low $n_g$, and possessing no
prominent preferred value at $n_g = 1$, contrary to what was seen in
the general population (see
Figure~\ref{fig:hist_ngbtr_Xray}). Similarly, the distribution in
bulge fraction was also found to show a deficit of $(B/T)_r < 0.3$ as
compared to the general population (see
Figure~\ref{fig:hist_ngbtr_Xray}). Taken together, therefore, the
structural parameter distributions seemed to indicate a bias towards
non-bulgeless galaxies in our X-ray selected sample.

\begin{figure}
\centerline{
\includegraphics[width=240pt,height=240pt,angle=0]{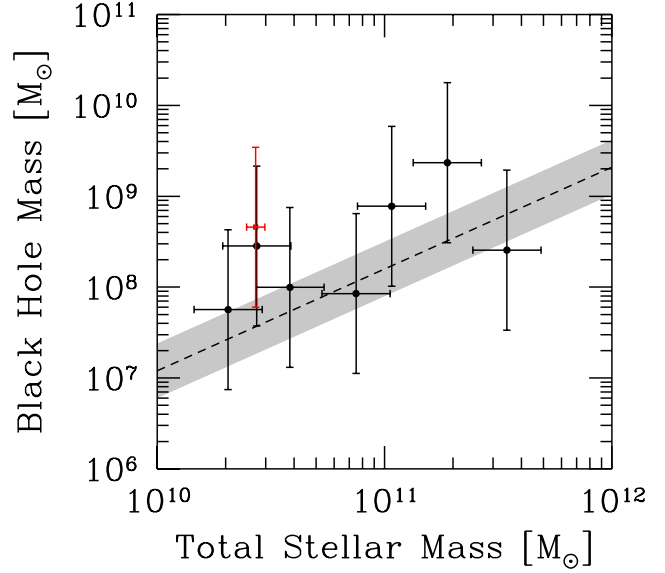}
}
\caption{\label{fig:bhmass_mstartot_Xray_err} Total stellar mass versus
  black hole mass estimated from the fundamental plane of BH 
  activity for our X-ray/radio sample ({\it black filled
    circles}). The bulgeless galaxy is shown as a {\it red filled
    square}. The horizontal error bars correspond to the 
    0.04 and 0.15 dex for the single- and two-component
    SED fits associated with the stellar mass estimates, 
    respectively (M12), while the vertical error bars 
    correspond to $\sigma_{R}=0.88$ of the fundamental plane
    of BH activity \citep{merloni03}. For comparison, 
    we also plot the empirical relation ({\it
    dashed line}) and the observed scatter of 0.3 dex ({\it grey
    shaded region}) of \citet{haring04}. \vspace{0.2cm}}
\end{figure}

\subsection{Stellar Mass Estimates}
\label{sec:xray.3}

One of the primary objectives of our work was to investigate the
relationship between the stellar masses of our galaxies and their
supermassive black holes. In order to obtain estimates of the stellar
masses of our sample, we made use of the catalog of \citet[hereafter
  \citetalias{mendel12}]{mendel12}. In this work,
\citetalias{mendel12} estimated the stellar masses of 669,634 galaxies
in the Legacy area of the SDSS DR7, with spectroscopic redshifts in
the range $0.005 \leq z \leq 0.4$.  Importantly for us,
\citetalias{mendel12} considered only sources classified as galaxies
both photometrically and spectroscopically by the SDSS, i.e.\ sources
with photoObj.type = 3 and specObj.specClass = 2. A key consequence of
the nature of this selection was that it should have had the effect of
filtering out most of the unobscured (type 1) AGN, i.e.\ those showing
broad emission lines. Crucially, this implies that our final sample
possessing stellar masses, obtained from the catalog of
\citetalias{mendel12}, both for this analysis and the analysis in
Section~\ref{sec:infrared}, should have been composed mainly of
obscured (type 2) AGN. Although, this might be viewed as a
restriction, as ideally one would like to have a sample composed of
both types of AGN, there are two immediate benefits to having a sample
composed of type 2 AGN. The first is that this provides an opportunity
to study SMBHs in type 2 AGN on a large scale.  The second benefit of
obtaining such a sample is that one can expect a minimal contribution
from obscured AGN to the total optical light in these galaxies, and
hence, a negligible effect from the AGN optical light to the estimates
of their stellar masses.

The stellar masses obtained by \citetalias{mendel12} were derived via
the fit of the spectral energy distribution (SED) of SDSS $ugri$
photometry with the flexible stellar population synthesis code of
\citet{conroy09}. For our selected sample of bulgeless galaxies, we
used the total stellar mass estimates from the S\'ersic profile
fitting \citepalias[Table~2]{mendel12}. While for the remainder of
our sample, we used the total, bulge and disk stellar mass estimates
from the de Vaucouleurs bulge + exponential disk fitting
\citepalias[Table~3]{mendel12}. It is important to note that the fit
to the total, bulge and disk photometry was carried out independently
and, therefore, it is not necessarily true that $M_{\star b} +
M_{\star d} \equiv M_{\star t}$. Based on Monte Carlo simulations, 
the uncertainties in stellar mass are estimated to be 0.04 and 0.15 dex 
for the single- and two-component fits, respectively \citepalias{mendel12}.
Of the 688 galaxies in our
X-ray/morphology sample, we were able to match 404 galaxies with the
\citetalias{mendel12} catalog and get their stellar masses.

\subsection{Black Hole Mass Estimates from Fundamental Plane of BH Activity}
\label{sec:xray.4}

To compute an estimate of the black hole mass for our X-ray sample, we
appealed to the fundamental plane of black hole activity
\citep{merloni03}, which correlates the X-ray and radio luminosities
with the mass of the black hole.

In order to search for the level of radio emission from the detected
Chandra X-ray sources in our galaxy sample, we utilized the source
catalog produced by the National Radio Observatory/Very Large Array
(NRAO/VLA) Sky Survey (NVSS). This radio survey was undertaken by the
NRAO VLA telescope and covered the sky north of a declination of -40
degrees, at a frequency of 1.4 GHz, a resolution of 45" and a limiting
peak source brightness of about 2.5 mJy/beam. Of the 404 sources that
we had previously identified as having been assigned a stellar mass in
the \citetalias{mendel12} catalog, we managed to find 8 that also had
an NVSS radio detection, one of which we identified as bulgeless.  The
stellar masses of our final 8 galaxies are given in
Table~\ref{tbl:tbl-1}, while their morphological parameters are listed
in Table~\ref{tbl:tbl-2}. We note that the galaxy 587724199889666262
can also be found in the 2MASS-selected Flat Galaxy Catalog of
\citet{mitronova04}.

The empirical relationship relating black hole mass to the emitted
compact radio and hard X-ray luminosities, spanning nine orders of
magnitude in black hole mass, is formalized by the equation
\begin{equation}
\log \, L_R = (0.60^{+0.11}_{-0.11}) \, \log L_X + (0.78^{+0.11}_{-0.09}) \, \log M + 7.33^{+4.05}_{-4.07},
\end{equation}
\noindent where $L_R$ is the radio luminosity at 5~GHz in erg/s, $L_X$
is the 2-10 keV X-ray luminosity in erg/s, and $M$ is the mass of the
black hole in solar masses. The scatter around this plane is 
$\sigma_R = 0.88$ \citep{merloni03}.

The observed 1.4~GHz radio flux densities were corrected to the 5~GHz
frequency by assuming a power-law emission of $S \propto \nu^{-0.7}$
at these frequencies, i.e.\ $S_{\rm 5GHz} = 0.41 \, S_{\rm
  1.4GHz}$. The correction to the rest-frame flux densities was
carried out by applying a multiplicative factor of $(1 +
z)^{0.7}$. The flux densities were also corrected for the effect of
bandwidth compression by applying the multiplicative term $(1 +
z)^{-1}$. The mass estimates of the black holes for our
final sample of 8 galaxies selected via their X-ray/radio emission are
given in Table~\ref{tbl:tbl-3}.

\begin{figure}
\centerline{
\includegraphics[width=240pt,height=240pt,angle=0]{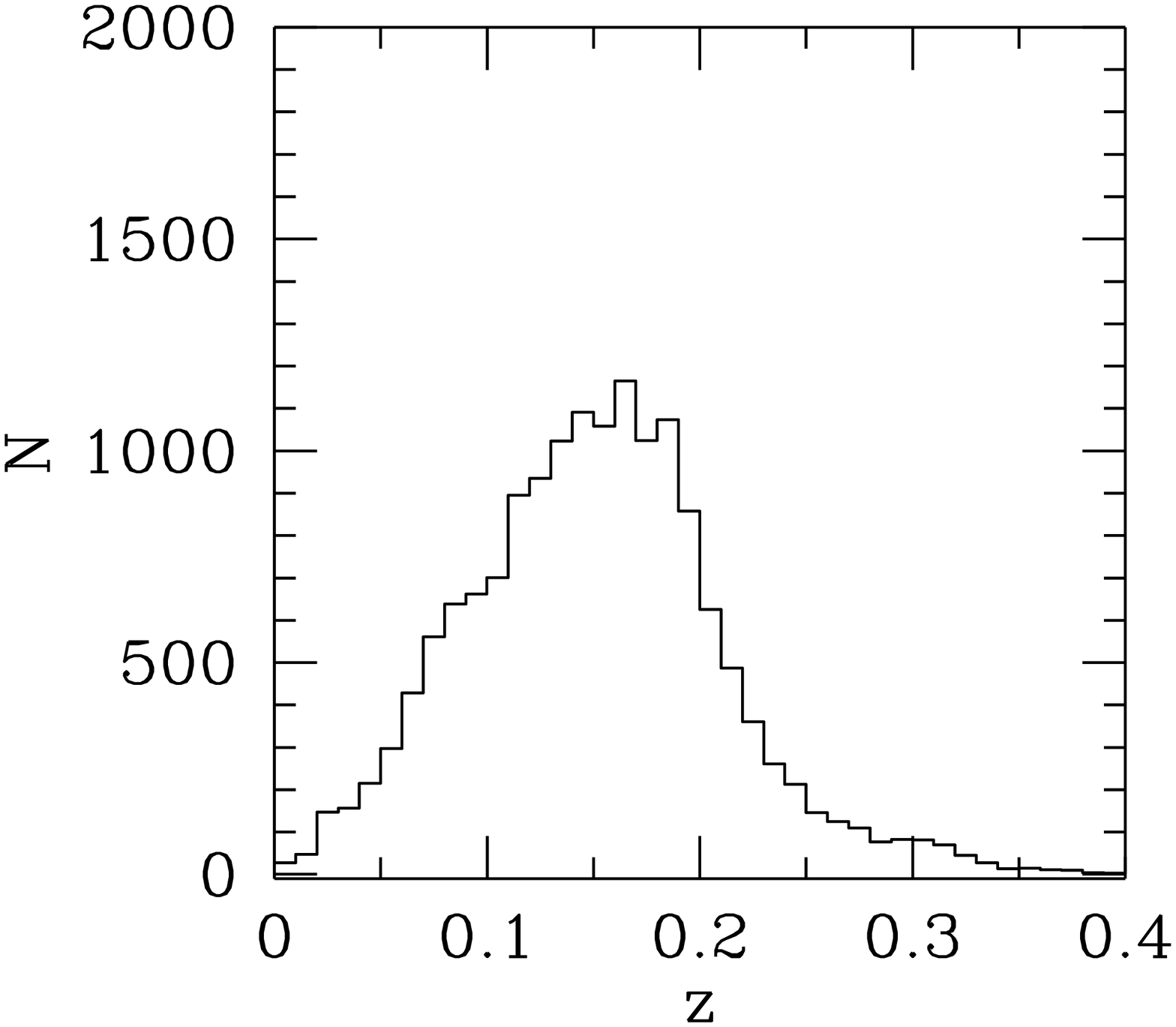}
}
\centerline{
\includegraphics[width=240pt,height=240pt,angle=0]{figures/hist_z_all.eps}
}
\caption{\label{fig:hist_z} {\it Top}: Redshift distribution of the 15,991 
 WISE color selected AGN with structural parameters. 
{\it Bottom}: Redshift distribution of the SDSS
  DR7 galaxies in the \citetalias{simard11} morphological catalog. \vspace{0.2cm}}
\end{figure}

\begin{table}
\caption{Stellar Mass Estimates \label{tbl:tbl-1}}
\scriptsize
\begin{center}
\begin{tabular}{crrr}
\hline
SDSS objid      &log(M$_{t}$/\msol) &log(M$_{b}$/\msol) &log(M$_{d}$/\msol)\\
\hline
587727229448421420    &10.3488  & 9.5920  &10.3382\\
588015508739195010    &11.2763  &11.2495  &10.3101\\
587731513142673426    &11.0313  &10.8544  &10.5517\\
587731513142673623    &10.4378  & 9.8747  &10.2581\\
587731513678561426    &10.3119  &10.2140  &10.2791\\
587731513687670913    &11.5380  &11.4924  &10.9015\\
587724199889666262    &10.5839  &10.4694  &10.4105\\
587730775500128413    &10.8746  &11.1970  & 8.6308\\
\hline
\end{tabular}
\end{center}
\end{table}

\begin{table}
\caption{Structural Parameter \label{tbl:tbl-2}}
\scriptsize
\begin{center}
\begin{tabular}{cccccc}
\hline
SDSS objid      &z &$n_g$ &$(B/T)_r$ &$n_b$ &$P_{pS}$\\
\hline
  587727229448421420  &0.048    &0.95    &0.00    &5.84   &0.500\\
  588015508739195010  &0.097    &4.97    &0.95    &5.77   &0.440\\
  587731513142673426  &0.045    &4.18    &0.65    &3.62   &0.210\\
  587731513142673623  &0.043    &1.28    &0.22    &3.13   &0.170\\
  587731513678561426  &0.094    &3.72    &0.94    &3.58   &0.500\\
  587731513687670913  &0.181    &4.66    &0.83    &4.03   &0.530\\
  587724199889666262  &0.041    &0.68    &0.28    &1.76   &0.020\\
  587730775500128413  &0.114    &3.00    &0.54    &6.34   &0.270\\
\hline
\end{tabular}
\end{center}
\end{table}

\begin{table}
\caption{Estimated Black Hole Mass \label{tbl:tbl-3}}
\scriptsize
\begin{center}
\begin{tabular}{cccc}
\hline
SDSS objid      &log($L_R$) &log($L_X$) &log($M$/\msol)\\
\hline
587727229448421420    &40.916  &38.634   &8.6598  \\
588015508739195010    &40.476  &38.924   &9.3690  \\
587731513142673426    &41.225  &38.999   &8.8898  \\
587731513142673623    &39.609  &37.689   &8.4525  \\
587731513678561426    &42.095  &38.634   &7.7521  \\
587731513687670913    &42.097  &39.146   &8.4072  \\
587724199889666262    &40.506  &37.871   &7.9969  \\
587730775500128413    &43.188  &39.427   &7.9291  \\
\hline
\end{tabular}
\end{center}
\end{table}

\begin{figure*}
\centerline{
\includegraphics[width=160pt,height=160pt,angle=0]{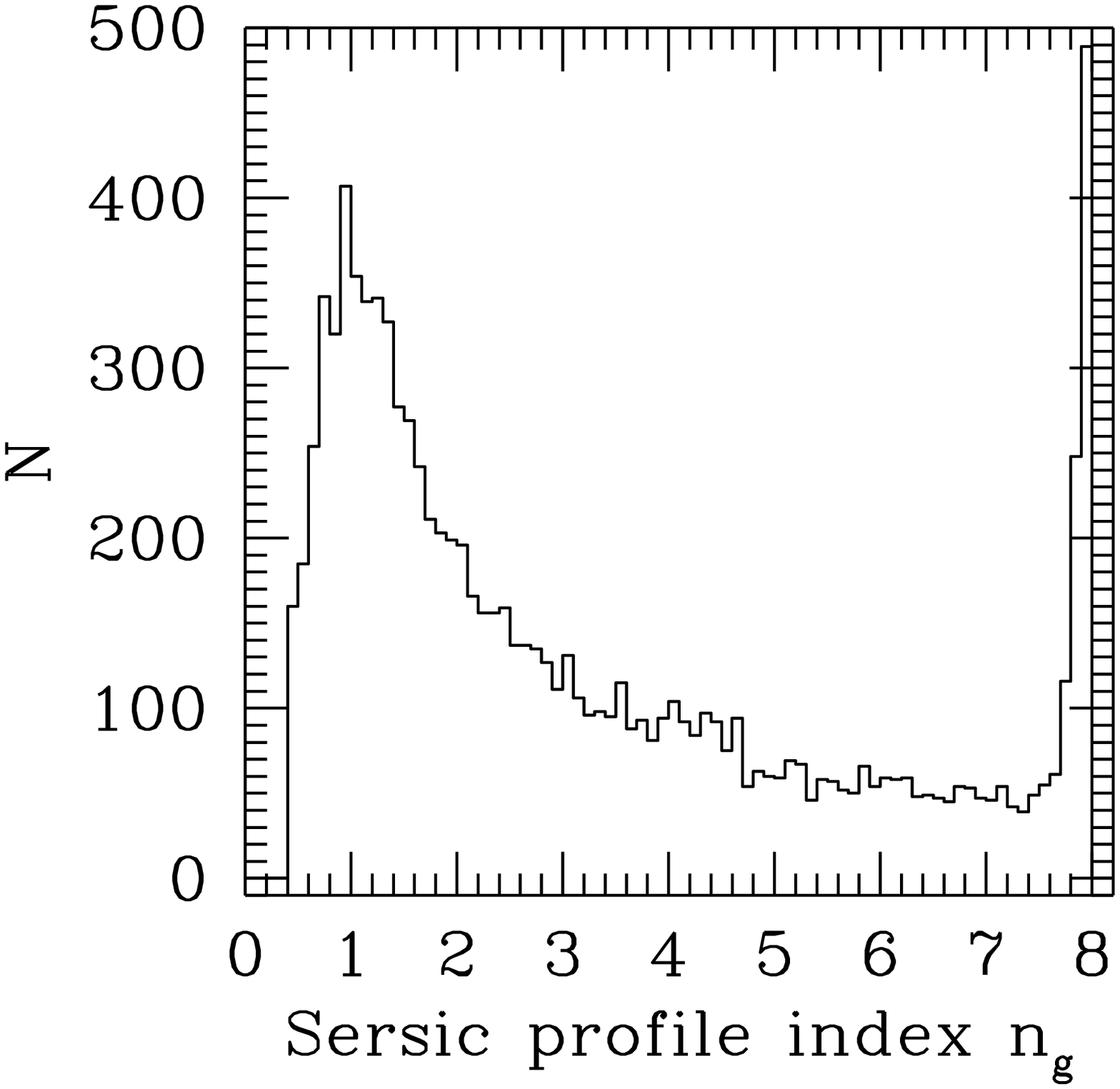}\includegraphics[width=160pt,height=160pt,angle=0]{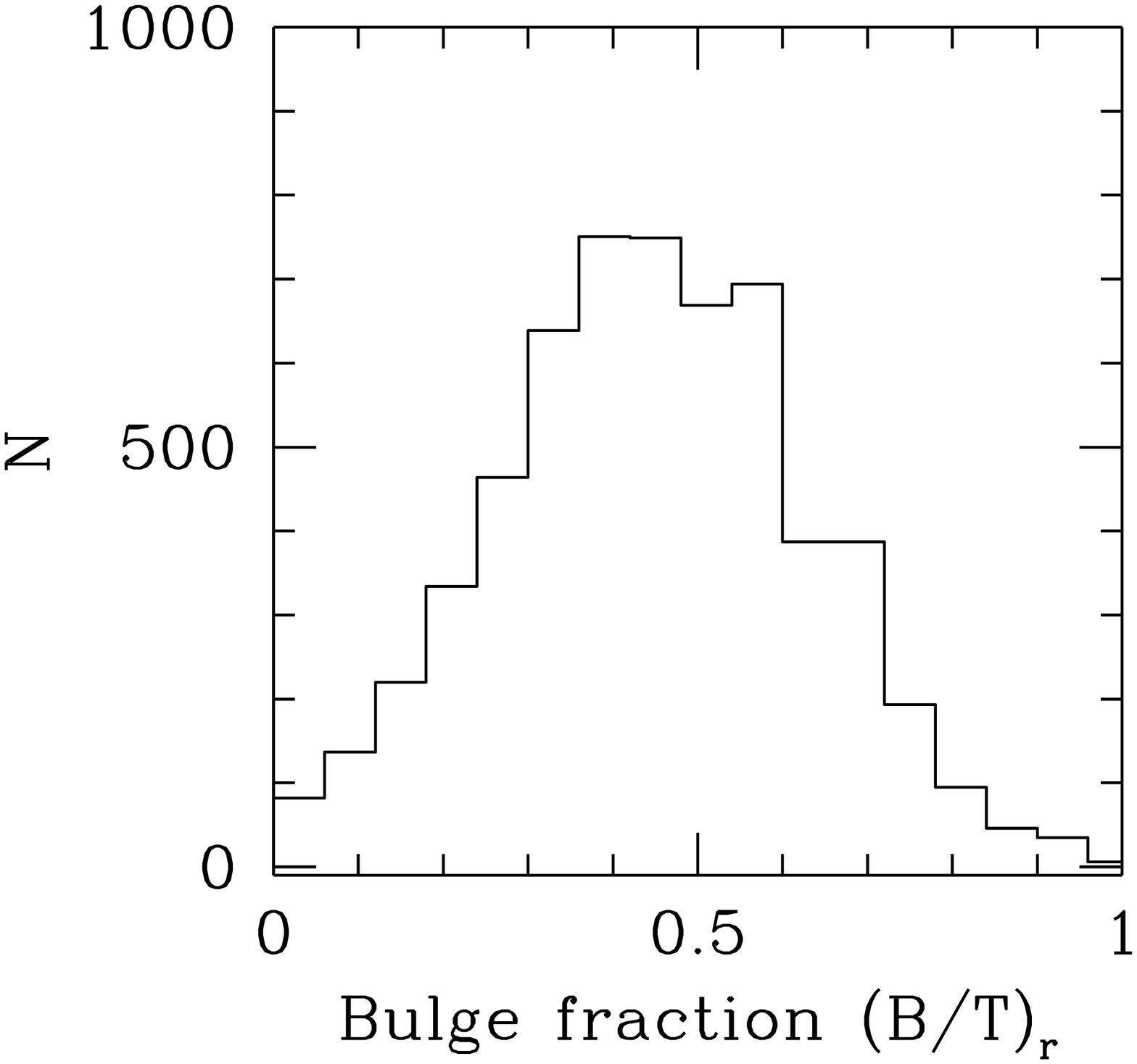}\includegraphics[width=160pt,height=160pt,angle=0]{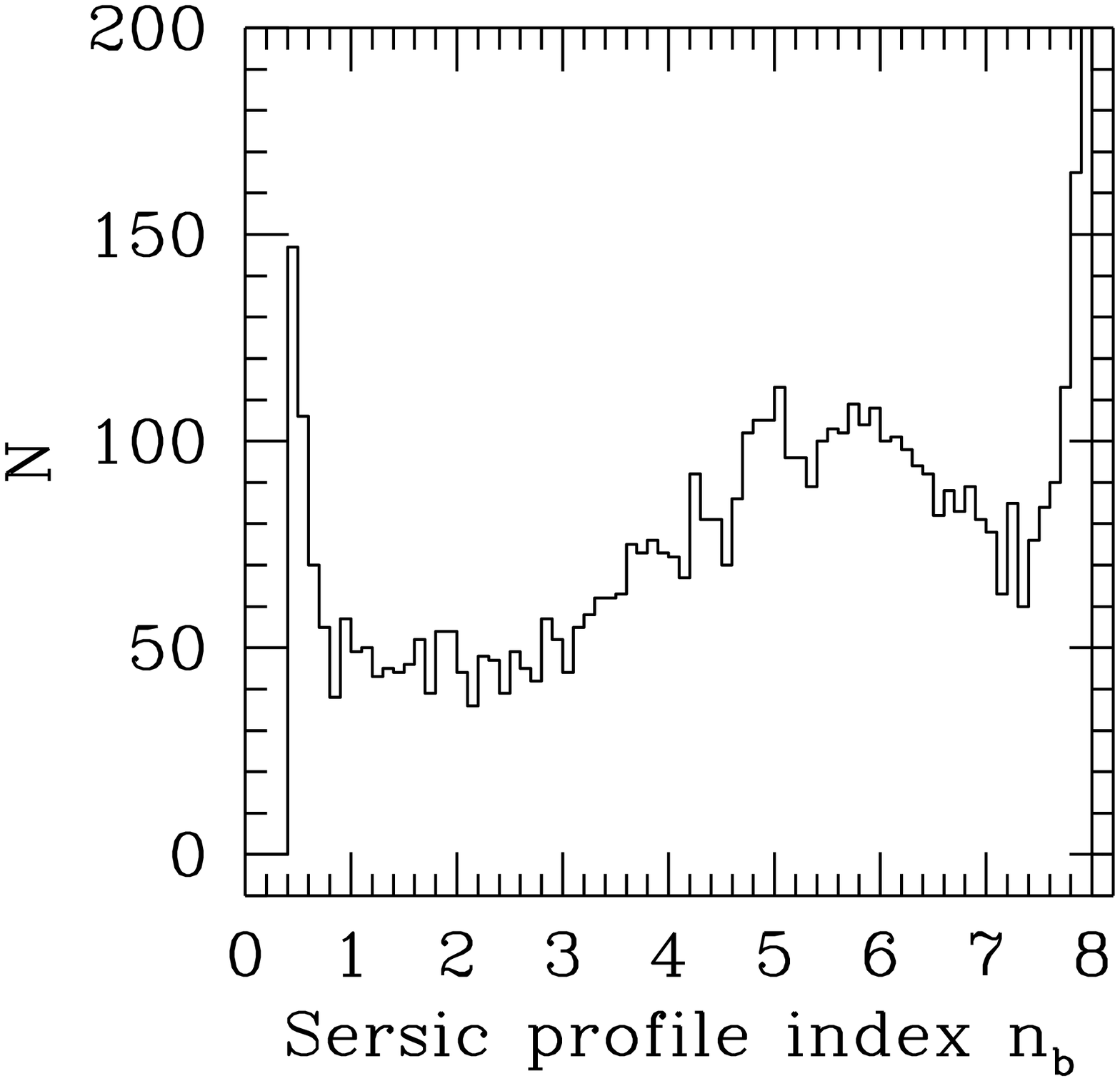}
}
\centerline{
\includegraphics[width=160pt,height=160pt,angle=0]{figures/hist_ng_all.eps}\includegraphics[width=160pt,height=160pt,angle=0]{figures/hist_btr_all.eps}\includegraphics[width=160pt,height=160pt,angle=0]{figures/hist_nb_all.eps}
}
\caption{\label{fig:hist_ngbtr} {\it Top row}: Distribution of the S\'ersic profile
  index $n_g$ ({\it left}), the bulge fraction $(B/T)_r$ ({\it
    middle}) and the bulge S\'ersic index $n_b$ ({\it right}) of the WISE
  color selected AGN with structural parameters. 
{\it Bottom row}: Distribution of the S\'ersic profile
  index $n_g$ ({\it left}), the bulge fraction $(B/T)_r$ ({\it
    middle}) and the bulge S\'ersic index $n_b$ ({\it right}) of the SDSS
  DR7 galaxies in the \citetalias{simard11} morphological catalog.
\vspace{0.2cm}}
\end{figure*}


The correlation between black hole mass and total stellar mass for our
sample of 8 galaxies, including the one classified as bulgeless, is
shown in Figure~\ref{fig:bhmass_mstartot_Xray_err}. For comparison, we
show in the same figure the empirical relation of \citet{haring04}
(dashed line) derived using dynamical and stellar masses of {\it
  bulge-dominated} systems, for which $M_{\star b} \simeq M_{\star
  t}$. In contrast, our {\it total stellar masses} are based either
on the S\'ersic profile fitting (for our bulgeless galaxy) or the de
Vaucouleurs bulge + exponential disk profile fitting (for the
remainder of the sample). The fact that the correlation between black
hole mass and total stellar mass is consistent with the relation of
\citet{haring04} suggests that the true correlation might not be with
the bulge mass but rather with the total stellar mass. That this is
indeed the case becomes clearer in our second study, which we now
detail.

\section{Infrared SMBH Sample Selection}
\label{sec:infrared}

\subsection{WISE-SDSS Cross-Match Catalog} 
\label{sec:infrared.1}

In our second analysis, we utilized the WISE all-sky catalog to
identify both unobscured (type 1) and obscured (type 2) AGN. In this
regard, it is important to note that most surveys of AGN that have
previously been undertaken have been biased towards observing mainly
unobscured (type 1) AGN, even though theoretical models generally
predict a large population of obscured (type 2) AGN, outnumbering type
1 AGN by a factor of $\sim$3 \citep[e.g.][]{comastri95, treister04,
  ballantyne11}.  A key advantage to using the mid-infrared selection
of AGN is that it does not suffer from this bias, as it relies upon
distinguishing the approximately power-law AGN spectrum from the black
body stellar spectrum of galaxies which peaks at rest-frame 1.6\mum,
and is thereby able to identify both unobscured (type 1) and obscured
(type 2) AGN.  Another advantage to working with mid-infrared data is
that they allow AGN to be easily distinguished from stars and
galaxies. Furthermore, an added benefit of mid-infrared selection is
that it does not suffer from dust extinction and is sensitive to the
highest redshift sources.

\begin{figure*}
\centerline{
\includegraphics[width=240pt,height=240pt,angle=0]{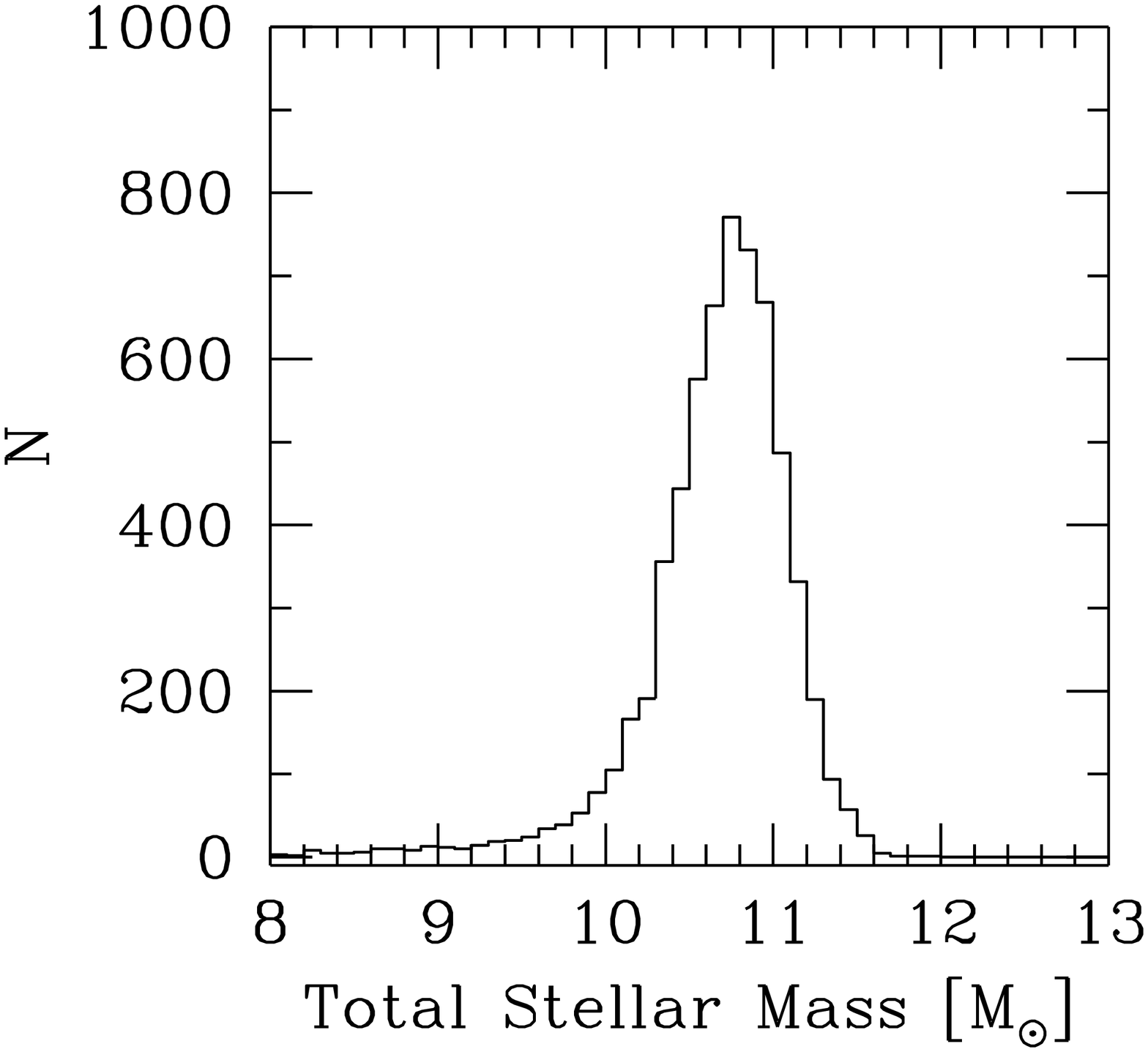}\includegraphics[width=240pt,height=240pt,angle=0]{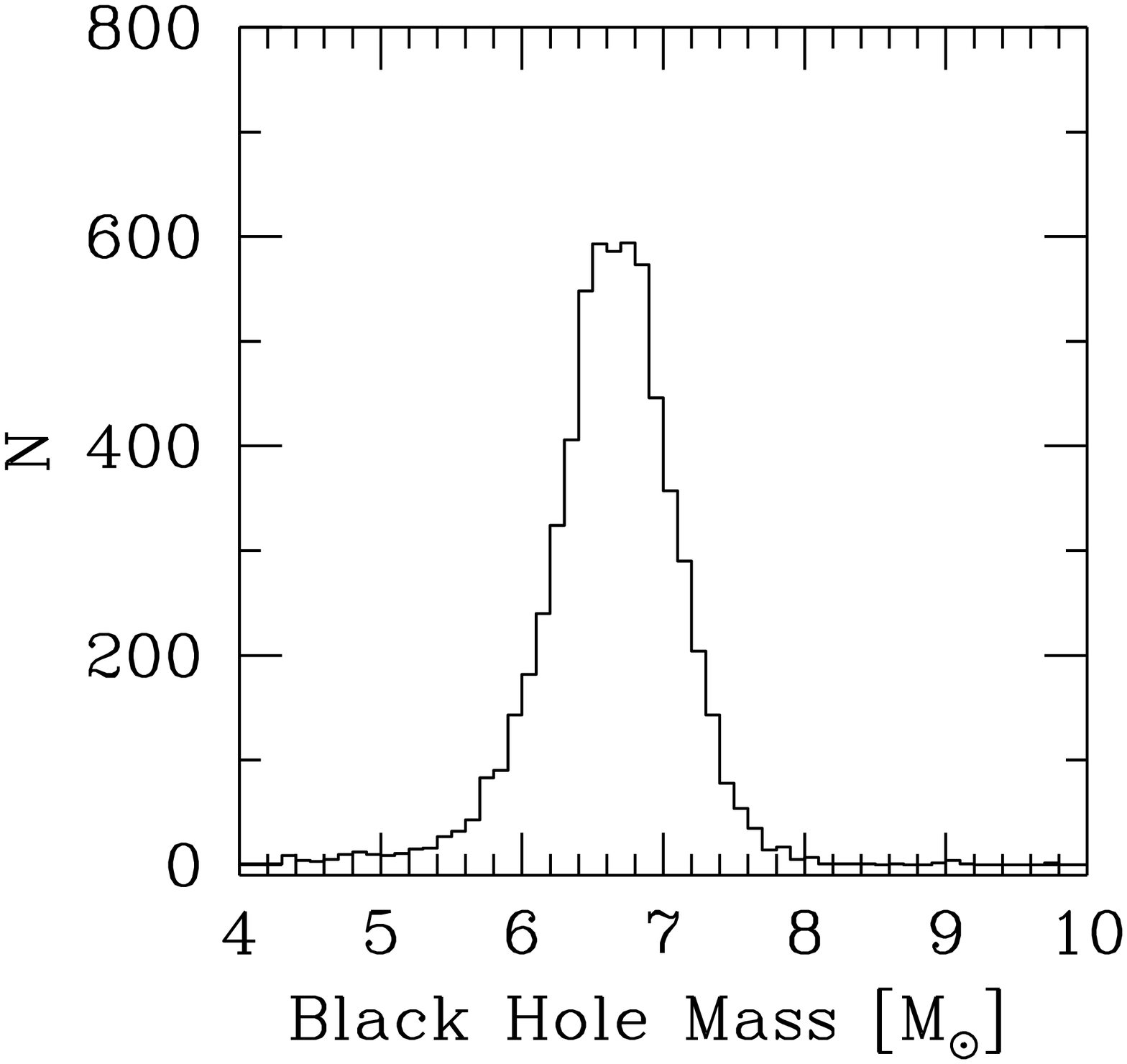}
}
\caption{\label{fig:hist_mstartotbhmass} Distribution of the total
  stellar masses ({\it left}) and lower limits on black hole masses
  ($L_{bol} = L_{Edd}$, {\it right}) of the 6,247 WISE color selected
  AGN with structural parameters and stellar masses. \vspace{0.2cm}}
\end{figure*}

In order to obtain our infrared sample we downloaded part22 to part50
of the WISE All-Sky Catalog (covering the area covered by SDSS,
i.e.\ dec $>$ -23.1899 degrees). The WISE All-Sky Release Source
Catalog contains positions and photometry at 3.4, 4.6, 12 and 22
micron for 563,921,584 point-like and resolved objects detected on the
Atlas Intensity images. Photometry was performed using point source
profile-fitting and multi-aperture photometry and the estimated
sensitivities are 0.068, 0.098, 0.86 and 5.4 mJy (5$\sigma$) at 3.4,
4.6, 12 and 22 micron in unconfused regions on the ecliptic
plane. J2000 positions and uncertainties were reconstructed using the
2MASS Point Source Catalog as astrometric reference. Astrometric
accuracy is approximately 0.2 arcsec root-mean-square on each axis
with respect to the 2MASS reference frame for sources with
signal-to-noise ratio greater than forty.

Based on their distinctive spectral energy distribution in the
infrared \citep{wu12, stern12, assef10}, we selected AGN in the WISE
all-sky catalog out to a redshift of $z \sim 3.5$ by applying the
mid-infrared color criterion $W1-W2 \geq 0.57$ \citep{stern12,
  wu12}. We found that $\sim$15\% of the WISE sources satisfied this
criterion (from a total of 3.27072 $\times 10^8$ sources, we were left
with 51,048,962.)

Having obtained our initial sample we then used the US Virtual
Astronomical Observatory (VAO) Cross-Comparison Tool to upload the
WISE-selected AGN catalog and cross-matched this with the online SDSS
DR 7 catalog, using a 0.5 arcsec match radius \citep[positional
  accuracy with 2MASS]{stern12}.  The number of cross-matched sources
equaled 6,969,533, corresponding to about 2\% of the initial sample.
In total, 15,991 galaxies were found to have structural parameters
derived by \citetalias{simard11} and we subsequently found that 1,450
of these galaxies (or about 9\%) satisfied our bulgeless selection
criteria.

\subsection{Redshift Distribution and Structural Parameters}
\label{sec:infrared.2}

The distributions of our WISE color selected AGN galaxy population
have been plotted in Figures~\ref{fig:hist_z} and
\ref{fig:hist_ngbtr}, as a function of their redshift, $z$, and
structural parameters, $n_g$, $(B/T)_r$ and $n_b$, respectively. As in
Figure~\ref{fig:hist_ngbtr_Xray} of Section~\ref{sec:xray.2}, the
S\'ersic profile index was plotted only for galaxies with $P_{pS} >
0.32$ \citepalias{simard11}, whereas both the bulge fraction and the
bulge S\'ersic index were plotted only for galaxies with $P_{pS} \leq
0.32$. Simple consideration of these distributions lead to some
important new results regarding the properties of AGN and, therefore,
more generally for galaxies containing SMBH.

\begin{figure*}
\centerline{
\includegraphics[width=240pt,height=240pt,angle=0]{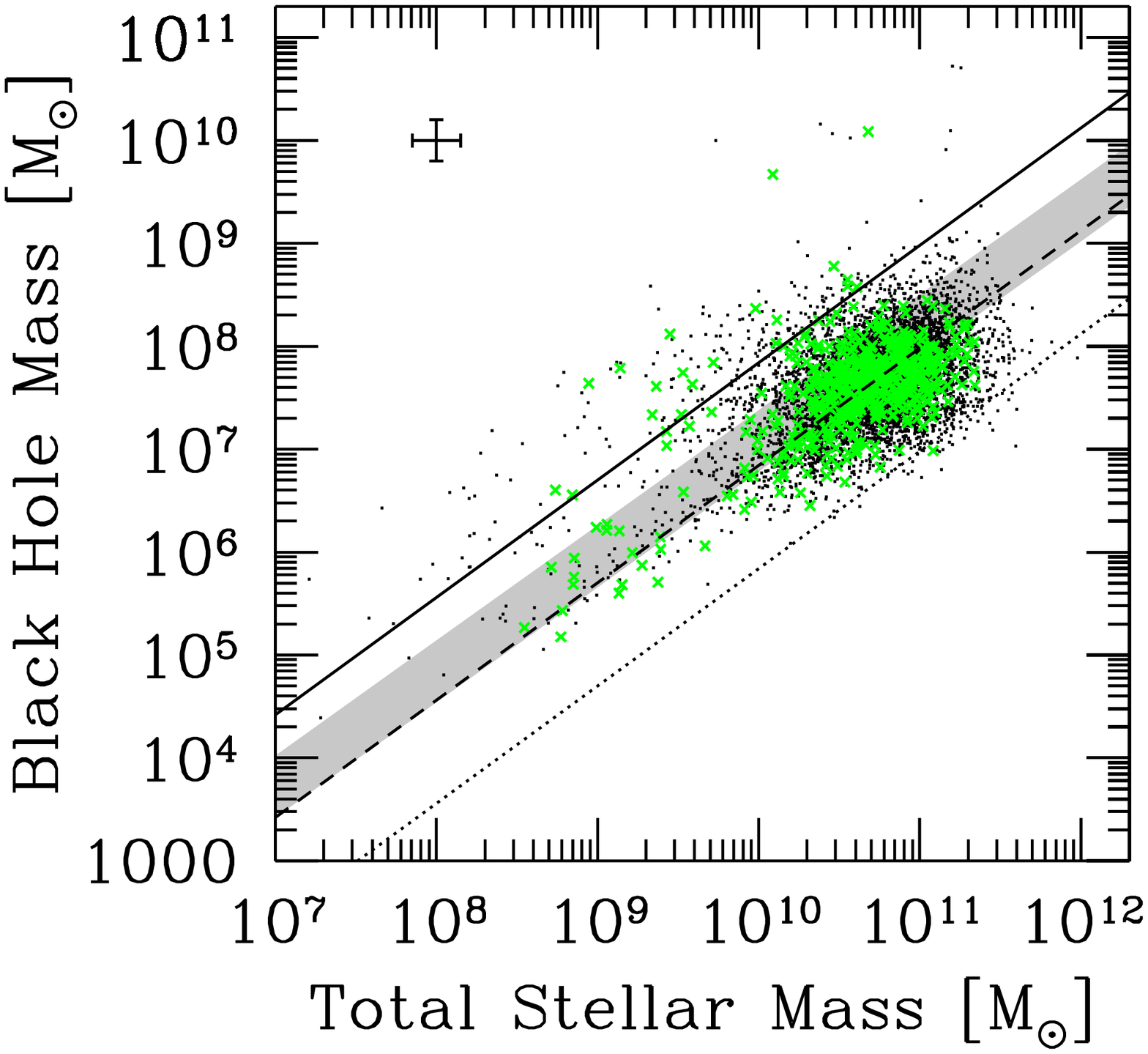}\includegraphics[width=240pt,height=240pt,angle=0]{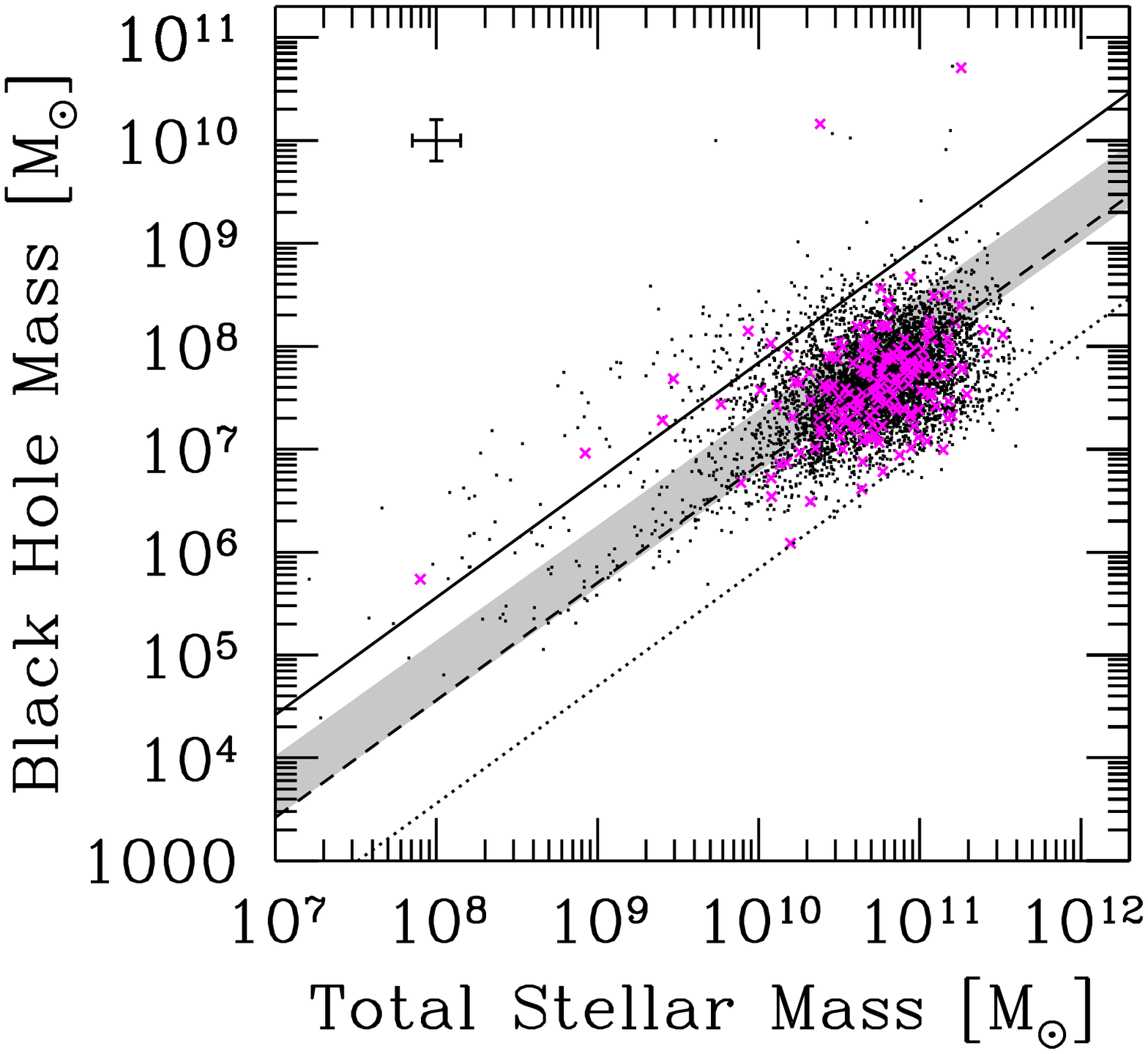}
}
\caption{\label{fig:bhmass_mstartot} {\it Left}: Total stellar mass
  versus black hole mass obtained using the bolometric luminosity for
  our infrared sample of 5,717 galaxies not classified as bulgeless
  ({\it black dots}) and for the 530 bulgeless galaxies ({\it
    green crosses}). The data points and bisector linear
  regression fit to the data ({\it dashed line}) are plotted for
  $L_{bol}/L_{Edd} = 0.1$. Also shown are the fit for $L_{bol}/L_{Edd}
  = 1.0$ ({\it dotted line}) and for $L_{bol}/L_{Edd} = 0.01$ ({\it
    solid line}). For comparison, we also plot the observed scatter of
  0.3 dex of the empirical relation of \citet{haring04} ({\it grey
    shaded region}). {\it Right}: Same as left figure, showing 
  in addition the galaxies with $3.9 < n_b < 4.1$ highlighted as 
  {\it magenta crosses}. \vspace{0.2cm}}
\end{figure*}

The first, and most important of these statistical results, which can
be immediately seen in the distribution of the S\'ersic index of our
infrared selected sample, Figure~\ref{fig:hist_ngbtr} (left), is that
the AGN are smoothly and significantly distributed over all values of
$n_g$. This necessarily implies that {\it AGN in our sample, and
  therefore their associated SMBH, are present in galaxies of all
  morphological types}. This is a key discovery, as until now
relatively few SMBH were known to exist in galaxies that did not
possess a significant bulge component.

The second important statistical result, is that the morphological
distributions of the AGN sample and the general population of galaxies
have a very similar form. This can be clearly seen by comparing the
S\'ersic distribution of the infrared selected AGN,
Figure~\ref{fig:hist_ngbtr} (top left), with that of the parent
population Figure~\ref{fig:hist_ngbtr} (bottom left). Alternatively,
comparing Figure~\ref{fig:hist_ngbtr} (top middle and right) with
Figure~\ref{fig:hist_ngbtr} (bottom middle and right), similarly
reveals that in the case when the galaxies have been fitted with a
bulge + disk component model, the galaxies with a SMBH again follow a
very similar distribution of bulge fraction and bulge S\'ersic index
to the general population of galaxies. To quantify this statement, we
used the Kolmogorov-Smirnov (K-S) test to measure the probability that
our first dataset is drawn from the same parent population as our
second dataset (the two-sample K-S test). The probability is a scalar
between 0 and 1 giving the significance level of the K-S statistic.
Small values of the K-S probability show that the cumulative
distribution function of the first dataset is significantly different
from the second dataset. The results of the K-S test reinforce our
qualitative comparison: after normalizing each distribution to the
total number count and using the same bin size as depicted in
Figure~\ref{fig:hist_ngbtr}, we compute K-S probabilities and maximum
deviations between datasets of (0.99,1.0,1.0) and (0.02,0.004,0.01)
for the distribution of ($n_g$,$(B/T)_r$,$n_b$), respectively,
indicating a very high probability that these three distributions are
the same.

Hence, from these comparisons we learn that not only are the SMBH in
our infrared selected sample distributed over all morphological galaxy
types, but that the form of their morphological distribution is very
similar to that of the general population of galaxies, over the same
redshift range, i.e.\ $0.05 < z < 0.25$. Moreover, this similarity
property is essentially robust with respect to the form of
morphological fitting that is used. An immediate consequence of this
approximate scaling relation between the distributions of the AGN and
the general population, is that it implies that {\it the fraction of
  galaxies that contain a SMBH is approximately the same for each
  morphological type}. Alternatively, one may say that the fraction of
a particular type of galaxy is the same in both the AGN and general
populations. In particular, the fraction of bulgeless galaxies that
host a SMBH is the same as the fraction of bulgeless galaxies in the
general population, which is $\sim$9\% in the redshift range
considered.

An important set of questions concern the extent to which our
statistical sample may be assumed to be representative of the entire
population of AGN galaxies and the general population of galaxies with
SMBH in the Universe. In the strictest sense the above results apply
only to our sample of infrared selected AGN. However, if our sample is
representative, which we reasonably expect it is, then it is also
reasonable to extrapolate the above results to the general population
of galaxies containing AGN. Furthermore, as the results of this
analysis suggest that AGN do not appear to significantly distinguish
between different galaxy morphologies, at least at the level we can
detect, it seems difficult to construct an argument to support the
idea that the more general population of galaxies with a SMBH
(i.e.\ active and inactive galactic nuclei) does. Moreover, if it is
the case that the above statistical results can be extrapolated to the
general population of galaxies containing a SMBH, they have an
important corollary. Since it follows that if the current widely held
opinion (and its supporting evidence) that most, if not all, massive
ellipticals and galaxies with bulges contain a SMBH is valid, then
essentially the same must apply to all the others, i.e.\ {\it most
  galaxies do contain a SMBH}.

\begin{figure*}
\centerline{
\includegraphics[width=240pt,height=240pt,angle=0]{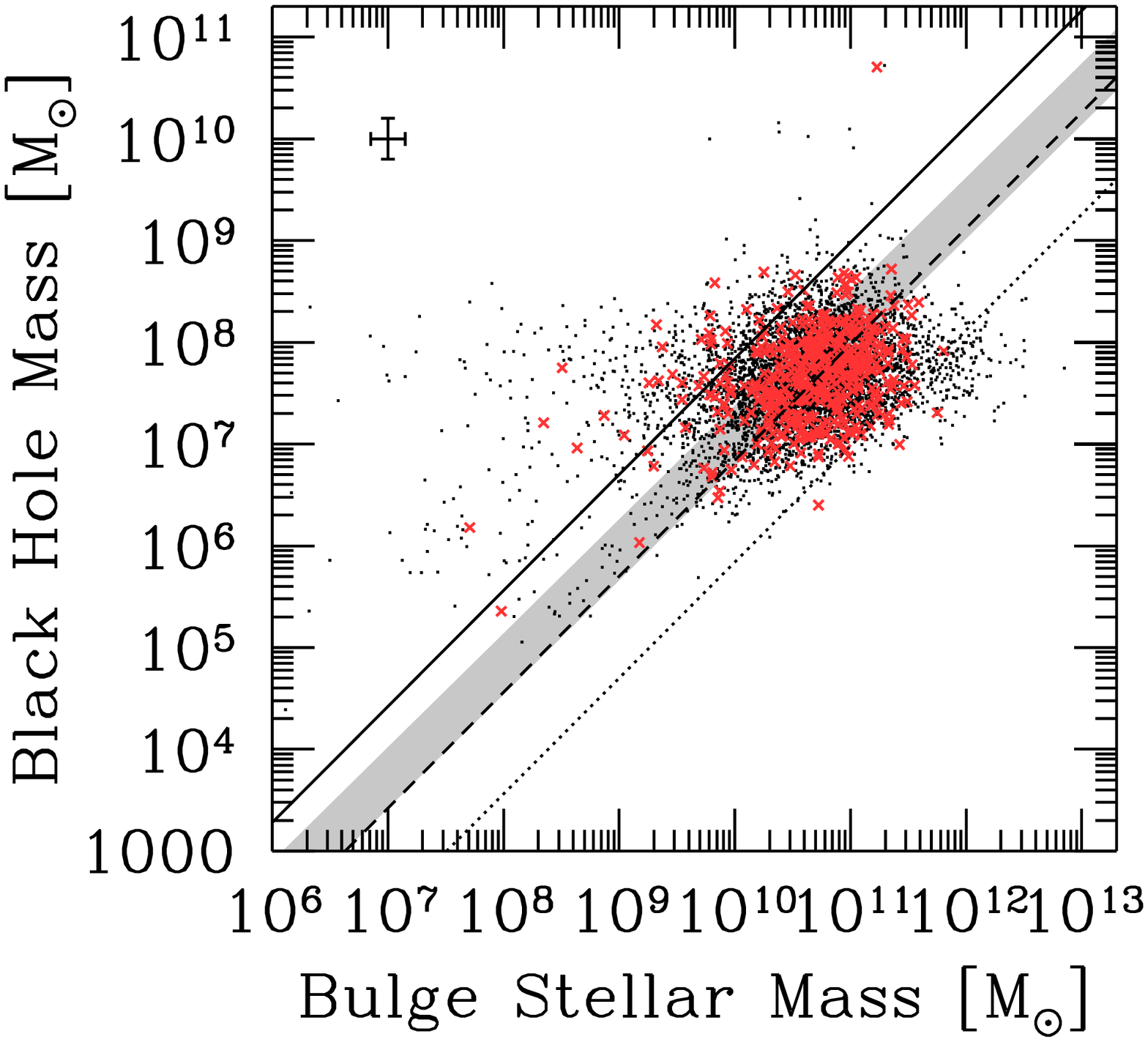}\includegraphics[width=240pt,height=240pt,angle=0]{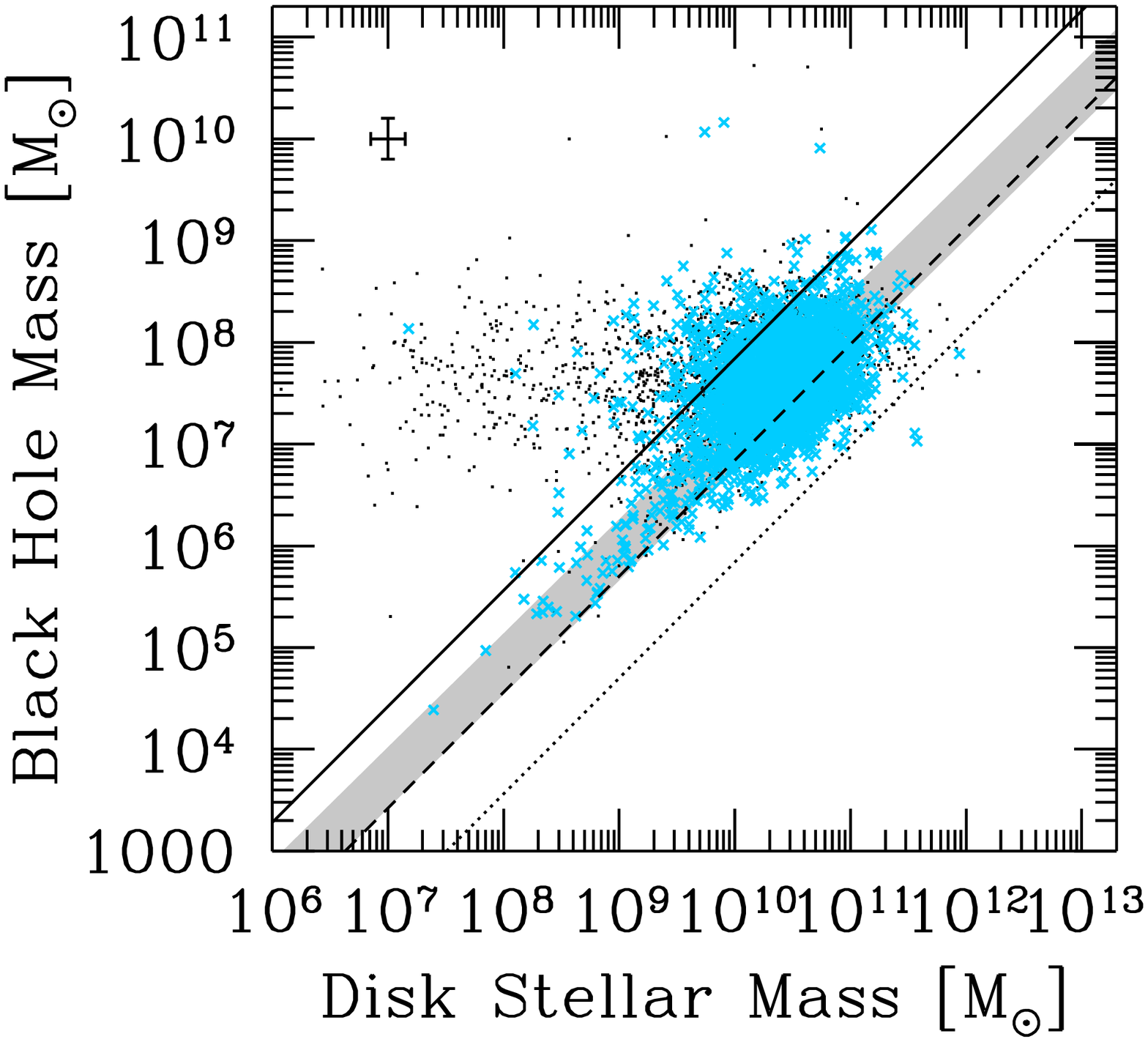}
}
\caption{\label{fig:bhmass_mstarbulgedisk} Bulge ({\it left}) and disk
  stellar mass ({\it right}) versus black hole mass obtained using the
  bolometric luminosity for our infrared sample of 5,717 galaxies not
  classified as bulgeless ({\it black dots}). The
  bulge-dominated galaxies only ($(B/T)_r \geq 0.5$) are highlighted
  as {\it red crosses} in the {\it left figure} and the
  disk-dominated galaxies only ($(B/T)_r < 0.5$) are highlighted as
  {\it blue crosses} in the {\it right figure}. As in
  Figure~\ref{fig:bhmass_mstartot}, the data points and bisector
  linear regression fit to the data ({\it dashed line}) are plotted
  for $L_{bol}/L_{Edd} = 0.1$. Also shown are the fit for
  $L_{bol}/L_{Edd} = 1.0$ ({\it dotted line}) and for $L_{bol}/L_{Edd}
  = 0.01$ ({\it solid line}). For comparison, we also plot the
  observed scatter of 0.3 dex of the empirical relation of
  \citet{haring04} ({\it grey shaded region}). \vspace{0.2cm}}
\end{figure*}

In comparing the redshift distribution of the WISE color selected AGN
with structural parameters with that of the SDSS DR7 galaxies in the
\citetalias{simard11} morphological catalog (see
Figure~\ref{fig:hist_z}), we noted a sharper fall-off toward lower
redshifts together with an associated shift in the peak of the
distribution from about $0.1$ to $0.2$.  This marked difference
between the two distributions originates from the removal of
apparently non-AGN galaxies from the sample at low redshifts by the
WISE cross-match selection process. One possible reason for this
apparent reduction of galaxies with AGN toward low redshifts could of
course be the absence or paucity of SMBH in more recently formed
galaxies. However, if this were the case, then as one would expect
newly formed galaxies to be biased toward particular morphology types,
one should similarly expect to see a difference between the
distribution of AGN in these morphology types with respect to the
general population, which as we have seen is not obviously
apparent. Hence, we are led to interpret the apparent depletion of AGN
toward low redshifts as being more likely primarily due to the
decrease in AGN activity toward low redshifts. Or, in other words, it
would seem that the difference between the two redshift distributions
should not be taken to imply that there exist fewer supermassive black
holes in galaxies at lower redshifts, but merely that black holes
become on average less active. This property, which is well known
\citep[e.g.][]{fan06}, can be seen for instance in the photometric
redshift distribution of the SDSS quasars and their WISE counterparts,
as depicted in Figure~2 of \citet{wu12}. Although here, by contrast,
we stress that we are witnessing this effect in both the type 1 and
type 2 AGN galaxy populations.


\subsection{Stellar Mass Estimates}
\label{sec:infrared.3}

Finally, we turn our attention to the relationship between the stellar
and black hole masses for our infrared selected sample of galaxies
with AGN. Following the description in Section~\ref{sec:xray.3} above,
we obtained the stellar masses of 6,247 galaxies in this sample from
\citet{mendel12}. As described in Section 3.3, the selection of sources
with stellar masses from \citet{mendel12} implied that we preferentially
selected obscured (type 2) AGN. In a similar fashion to our X-ray/radio sample, we
used the S\'ersic profile total stellar mass estimates for our 530
bulgeless galaxies and the de Vaucouleurs + exponential profile total,
bulge and disk stellar mass estimates for the remainder of our
sample. The distribution of total stellar masses is shown in
Figure~\ref{fig:hist_mstartotbhmass} (left). For clarity, the
sizes of the statistical samples used in our analysis are summarized
in Table~\ref{tbl:tbl-4}.

\begin{table}
\caption{Sizes of Statistical Samples \label{tbl:tbl-4}}
\scriptsize
\begin{center}
\begin{tabular}{lrr}
\hline
Sample                                                  &No.\ Galaxies           &No.\ Bulgeless\\
\hline
S11 catalogue:                                          &1,123,718               &155,040 \\
S11 + $0.05 < z < 0.25$:                                & 946,299                &110,913 \\
WISE catalogue:                                         &3.27072 $\times 10^8$   &N/A \\
WISE + $W1-W2 \geq 0.57^a$:                             &51,048,962              &N/A \\
WISEcol + SDSS:                                         &6,969,533               &N/A \\
WISEcol + SDSS + S11:                                   &15,991                  &1,450 \\
WISEcol + SDSS + S11 + M12:                             &6,247                   &530 \\
CSC-SDSS catalogue:                                     &8,997                   &N/A \\
CSC-SDSS + NVSS:                                        &34                      &N/A \\
CSC-SDSS + S11:                                         &688                     &26 \\
CSC-SDSS + NVSS + S11:                                  &10                      &1 \\
CSC-SDSS + NVSS + S11 + M12:                            &8                       &1 \\
\hline
\end{tabular}
\end{center}
$^a$hereafter WISEcol.
\end{table}

\subsection{Black Hole Mass Estimates from the Bolometric Luminosity}
\label{sec:infrared.4}

Black hole masses were estimated using the bolometric luminosity of
the galaxies. Bolometric luminosities \citep[taken to be the 100
  micron to 10 keV integrated luminosity][]{richards06} are typically
obtained using corrections to the mid-infrared bands where the AGN
emission dominates. We used the 12 (W3) and 22 micron (W4) k-corrected
flux densities from WISE to compute the bolometric luminosities of our
WISE color-selected AGN and applied the bolometric corrections
$L_{bol} \simeq 8 \times L_{12\mu m}$ (W3) and $L_{bol} \simeq 10
\times L_{22\mu m}$ (W4) from \citet{richards06}, which are not
strongly dependent on AGN luminosity (see their Figure~12). Given the
bolometric luminosity, making the assumption that accretion is at the
Eddington limit yields a lower limit on the black hole mass. However,
given that the growth rates of AGN can span several orders of
magnitude, from super-Eddington accretion to $10^{-3} L_{Edd}$
\citep{simmons13, steinhardt10}, we computed black hole masses
assuming the following three cases: $L_{bol} = L_{Edd}$, $L_{bol} =
0.1 L_{Edd}$, and $L_{bol} = 0.01 L_{Edd}$. The histogram of BH masses
($L_{bol} = L_{Edd}$) is shown in Figure~\ref{fig:hist_mstartotbhmass}
(right). The black hole masses range from $10^5$\msol\ to
$10^8$\msol\ with a peak near $\sim 4 \times 10^{6}$. These numbers
are consistent with those measured in the sample of \citet{simmons13}
($10^5$\msol\ to $10^6$\msol) given our larger sample size.  However,
they are statistically lower than the masses measured from our
X-ray/radio sample.

Using the stellar masses of \citet{mendel12} and the black hole masses
estimated from the bolometric luminosity of the galaxies, we present
the correlation between black hole mass and total stellar mass for our
sample of 6,247 galaxies, including the 530 classified as bulgeless in
Figure~\ref{fig:bhmass_mstartot} (left). Here, we only show the
results based on the 22 micron luminosities, although the results
based on the 12 micron luminosities are very similar.

The first thing to note is that the results of using our new method of
selecting AGN based on their infrared colors clearly confirm the
strong correlation between the total stellar mass of galaxies and the
mass of their SMBHs. However importantly, whereas previous studies
were limited to primarily bulge-dominated systems, our study confirms
this relationship to {\it all morphological types, in particular, to
  bulgeless galaxies} (Figure~\ref{fig:bhmass_mstartot} left, green
crosses).  Additionally, we emphasize that this study is by far the
largest sample to date.

In this figure, the total stellar mass estimates for the bulgeless
galaxies are from the S\'ersic profile fitting
(Figure~\ref{fig:bhmass_mstartot} left, green crosses) whereas the
rest are from the de Vaucouleurs bulge + exponential disk fitting
(Figure~\ref{fig:bhmass_mstartot} left, black dots). For reference to
previous work done for classical bulges, we show in the same figure
the sample of galaxies with bulge S\'ersic index $3.9 < n_b < 4.1$
(Figure~\ref{fig:bhmass_mstartot} right, magenta crosses). It may
be noted that the location of these galaxies in the diagram is almost
indistinguishable from the bulgeless galaxies.

A bisector linear regression fit to the data leads to the relation:

\begin{equation}
\log \, \left(\frac{M_{bh}}{\msol}\right) = \alpha + \beta \, \log \left(\frac{M_{\star t}}{10^{11} \msol}\right),
\end{equation}

where $\alpha = 7.98 \pm 0.20$ and $\beta = 1.14 \pm 0.10$ for
$L_{bol}/L_{Edd} = 0.1$. This relation is plotted in
Figure~\ref{fig:bhmass_mstartot} with a dashed line. The relations
obtained using $L_{bol}/L_{Edd} = 1.0$ and $L_{bol}/L_{Edd} = 0.01$
are also shown for comparison with a dotted and solid line,
respectively.

For comparison, we have also shown in the same figure the empirical
relation of \citet{haring04} (where $\alpha = 8.20 \pm 0.10$, $\beta =
1.12 \pm 0.06$, and scatter of 0.3 dex, grey shaded region). Although
the slopes of the two fits are essentially the same, the agreement on
the intercept depends on the value of $L_{bol}/L_{Edd}$. Assuming that
previous studies such as that of \citet{haring04} based on local
dynamical mass estimates are correct, then our results appear to be
most consistent with the black hole masses derived using
$L_{bol}/L_{Edd} = 0.1$. In fact, we expect this assumption to be the
main source of uncertainty in our analysis, as the errors that arise
from estimating the total stellar masses are on average $\sim$0.15 dex
and those from estimating the bolometric luminosity are on average
$\sim$0.20 dex, and hence cannot account for such a shift. Indeed, in
this regard, \citet{simmons13} point out that accretion may take place
at well below the Eddington limit and find that, at least for two
sources in their sample, $L_{bol}/L_{Edd} = 0.05$ and 0.08.  Hence, if
a growth rate of order $L_{bol}/L_{Edd} = 0.1$ is typical of the
sample, it implies that our black hole and total stellar mass relation
is consistent with the empirical relation of \citet{haring04}.

In Figure~\ref{fig:bhmass_mstarbulgedisk}, we have plotted the results
which essentially represent the galaxies in our sample with a
significant bulge component, specifically those that do not satisfy
the bulgeless selection criteria $n_g < 1.5$ and $(B/T)_r < 0.1$. In
Figure~\ref{fig:bhmass_mstarbulgedisk} (left) we consider the bulge
stellar mass only, while in Figure~\ref{fig:bhmass_mstarbulgedisk}
(right) we consider the disk stellar mass only. In both figures
scatter is significantly increased with respect to
Figure~\ref{fig:bhmass_mstartot}, which uses the total stellar
mass. This indicates that {\it the true correlation is between the
  black hole mass and the total stellar mass}. This is confirmed when
we consider the bulge-dominated galaxies only $(B/T)_r \geq 0.5$) in
Figure~\ref{fig:bhmass_mstarbulgedisk} (left, red crosses), and the
disk-dominated galaxies only $(B/T)_r < 0.5$) (right, blue crosses),
and note that these exhibit significantly tighter correlations.
Therefore, we conclude that the previous assumption that the black
hole mass was correlated with the bulge mass was only approximately
correct, since the true relation is in fact with the total stellar
mass of the galaxy. However, the approximate bulge mass correlation
therefore becomes increasingly accurate as the bulge mass dominates
the total mass of the galaxy.

\begin{figure}
\centerline{
\includegraphics[width=240pt,height=240pt,angle=0]{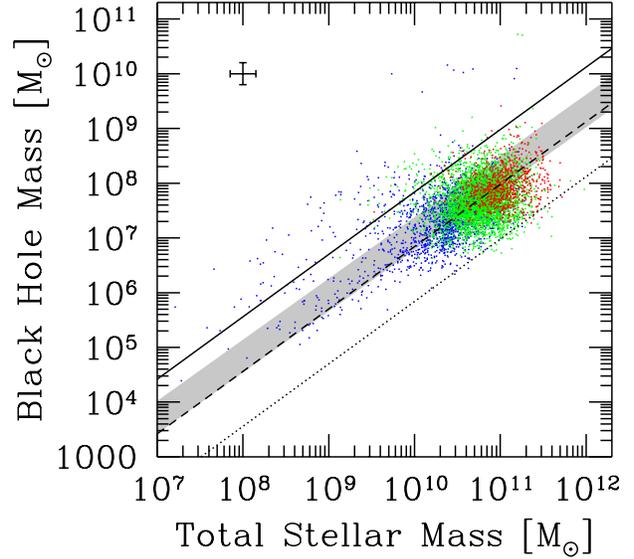}
}
\caption{\label{fig:bhmass_mstartot_z} Same as Figure~\ref{fig:bhmass_mstartot},
with galaxies divided in three redshift bins: $z < 0.1$ ({\it blue}), 
$0.1 \leq z < 0.2$ ({\it green}), and $z \ge 0.2$ ({\it red}). \vspace{0.2cm}}
\end{figure}

In Figure~\ref{fig:bhmass_mstartot_z}, we also provide a plot of the
data sorted into three redshift bins: $z < 0.1$ (blue open circles),
$0.1 \leq z < 0.2$ (green open circles), and $z \ge 0.2$ (red open
circles).  Comparing data at low redshift, it is entirely consistent
with what is found based on local dynamical mass estimates. In this
figure, there appears to be a deficit of low masses at high redshift.
This redshift dependence is very likely due to the bias introduced by the
limiting magnitudes of both the SDSS and WISE all-sky survey and 
not to any evolutionary effect. For instance, the minimum logarithm 
of the BH mass based on the flux limit of WISE at 22 micron of 5.4 mJy 
(5$\sigma$) and using $L_{bol} = 0.1 \; L_{Edd}$ is
(5.0,7.0,7.7,8.0) at z=(0.01,0.1,0.2,0.3), respectively.

\section{Summary}
\label{sec:summary}

By using the quantitative morphological analysis of 1.12 million
galaxies in the SDSS DR7 \citepalias{simard11}, we have presented the
first systematic survey of SMBHs in galaxies of all morphological
types, i.e.\ including bulgeless galaxies. Two different methods were
used to identify the presence of black holes via their activity, the
first being their X-ray/radio emission and the second their
mid-infrared colors. These two studies are complementary in that while
our first analysis of the X-ray selected sample is biased towards
selecting unobscured AGN and provides a well-tested and robust
estimate of black hole mass but with a small statistic, our second
analysis of the infrared selected sample includes both unobscured
(type 1) and obscured (type 2) AGN and provides estimates on the black
hole mass but with a large statistic.

Our first analysis identified 688 unobscured active galaxies with
known structural parameters. When the S\'ersic profile index $n_g$ and
bulge fraction $(B/T)_r$ of these galaxies were examined, we were able
to identify 26 of these galaxies hosting a SMBH as bulgeless. We were
able to obtain estimates for both the stellar and black hole masses
for 8 of these 688 galaxies and investigate their relationship. The
results of this first analysis confirm the correlation between black
hole and total stellar mass for these 8 galaxies and includes one
galaxy classified as bulgeless.

Our second analysis identified 15,991 mostly obscured active galaxies
with known structural parameters. When the S\'ersic profile index and
bulge fraction of this sample were examined, we were able to identify
1,450 bulgeless galaxies containing AGN. Moreover, when analyzing the
full range of structural parameters, we found that these active
galaxies followed a very similar morphological distribution to the
general population of galaxies in the same redshift range. In
particular, the fraction of bulgeless galaxies, $\sim$ 9\%, was found
to be the same as in the general population. For this larger sample,
we were also able to obtain estimates of the total stellar and black
hole masses for 6,247 galaxies and investigate their
relationship. Importantly, whereas previous studies were limited to
primarily bulge-dominated systems, our study confirms this
relationship to all morphological types, in particular, to 530
bulgeless galaxies. Furthermore, we were able to investigate the same
relationship using the independently estimated bulge and disk stellar
masses. Our results indicated that the true correlation is between the
black hole mass and the total stellar mass and hence we concluded that
the previous assumption that the black hole mass is correlated with
the bulge mass was only approximately correct. If extrapolated, the
combination of these results appear to indicate that SMBHs are generic
to galaxies regardless of the details of stellar galaxy evolution,
dynamics and merger history of galaxies. As it is only possible to
estimate dynamical masses locally, the future of studying SMBHs in
galaxies at all redshifts lies in a method such as the one presented
in this work.

\section{Future Work}
\label{sec:future}

An obvious extension of our analysis, to be presented in a future
paper, will be to include unobscured AGNs. For these galaxies, where
the nuclear emission is clearly visible, fitting an additional
component to the two-dimensional light profile will be necessary, as
the fitting models used in \citetalias{simard11} would underestimate
the bulgeless fraction of the AGN sample by overestimating the
S\'ersic index of the host component.

\end{document}